\documentclass[12pt,dvips]{article}
\usepackage{rotating}
\usepackage{graphicx,epsfig}
\usepackage{color}
\usepackage{cite}

\newcommand{\lsim}{\raisebox{-0.13cm}{~\shortstack{$<$ \\[-0.07cm] $\sim$}}~}
\newcommand{\gsim}{\raisebox{-0.13cm}{~\shortstack{$>$ \\[-0.07cm] $\sim$}}~}
\newcommand{\psl}{\mathbf{p} \hspace{-0.5 em}/}

\begin{document}

\def\thefootnote{\fnsymbol{footnote}}

\begin{flushright}
\today
\end{flushright}

\begin{center}
{\bf {\LARGE
Reconstructing the Higgs boson \\[2mm] in dileptonic $W$ decays at
hadron collider
} }
\end{center}

\medskip

\begin{center}{\large
Kiwoon~Choi$^a$, Suyong~Choi$^b$, Jae~Sik~Lee$^c$,
Chan~Beom~Park$^a$ }
\end{center}

\begin{center}
{\em $^a$Department of Physics, KAIST, Daejeon 305-701, Korea}\\[0.2cm]
{\em $^b$Sungkyunkwan University, Suwon 440-746, Korea}\\[0.2cm]
{\em $^c$Physics Division, National Center for Theoretical Sciences,
Hsinchu, Taiwan}\\[0.2cm]
\end{center}

\bigskip

\centerline{\bf ABSTRACT}
\begin{flushleft}
\medskip\noindent
We examine the prospect to measure the Higgs boson mass using the
recently introduced kinematic variable, the $M_{T2}$-Assisted
On-Shell (MAOS) momentum, that provides a systematic approximation
to the invisible neutrino momenta in dileptonic decays of $W$-boson
pair. For this purpose, we introduce a modified version of the MAOS
momentum, that is applicable even when one or both of the $W$-bosons
from  the Higgs decay
are in off-shell. It is demonstrated that the MAOS Higgs mass
distribution, constructed with the MAOS neutrino momenta, shows a
clear peak at the true Higgs boson mass when an event cut selecting
higher value of $M_{T2}$ is employed.
We perform the likelihood analysis for this MAOS  mass distribution
to determine the Higgs boson mass, and find it can improve the
accuracy of the Higgs mass measurement. Our results indicate that
the MAOS Higgs mass can be useful also for the discovery or
exclusion of the Higgs boson in certain mass range.

\end{flushleft}

\newpage

\section{Introduction}

Hunting down the Higgs boson, the last ingredient of the Standard Model (SM),
is one of the most important tasks of the LHC~\cite{Aad:2009wy,Ball:2007zza}.
The LEP experiments established a lower bound of 114.4 GeV, at 95\%
confidence level, on the mass of the SM Higgs
boson~\cite{Barate:2003sz}.
On the other hand, the electroweak precision data points towards a
relatively light SM Higgs boson with $m_H \stackrel{<}{{}_\sim}
185$~GeV  at 95\% confidence level~\cite{Collaboration:2008ub}.
Recently, Tevatron data have excluded the SM Higgs mass in the range
160 GeV $\leq m_H\leq$ 170 GeV again at 95\% confidence level
\cite{Collaboration:2009pt}.

The strategy for the Higgs boson search  depends on its decay
pattern. The SM Higgs boson lighter than about $180$ GeV mainly
decays into the $b$ quarks and $W$ bosons. At  hadron collider such
as the Tevatron or the LHC, Higgs boson search through the $b$ quark
channel appears to be very difficult due to overwhelming  QCD
backgrounds.
%
In this respect, the Higgs decay  $H \to WW \to l\nu \,
l^\prime\nu^{\,\prime}$ with $l\,,l^\prime =e\,,\mu$ may provide the
best search channel  for the SM Higgs boson in the mass range $135
\,\,{\rm GeV}\leq m_H\leq 180 \,\, {\rm GeV}$. Even for a heavier
Higgs boson with $m_H \gsim 2 M_Z$, this channel gets benefit from a
larger branching ratio compared to the decay
 $H \to ZZ \to 4\,l$. A drawback is that this channel involves
two invisible neutrinos, making it impossible to reconstruct the
Higgs boson mass directly. One then has to rely on a Higgs-induced
excess in the distribution of certain observables, and this
procedure typically requires an accurate estimate of the background
contributions \cite{Buescher:2005re}.

Recently, a new collider  variable,  the $M_{T2}$-Assisted On-Shell
(MAOS) momentum, has been introduced \cite{Cho:2008tj} to
approximate the invisible particle momenta in the process
$XX^\prime\rightarrow V\chi V^\prime\chi^\prime,$ where $X$ and
$X^\prime$ denote pair-produced mother particles, $V$ and $V^\prime$
represent the visible particles (one or more particles for each of
them) produced by the decays of $X$ and $X^\prime$, respectively,
and $\chi$ and $\chi^\prime$ are the invisible particles having the
same  mass. In this paper, we examine the possibility to determine
the Higgs boson mass using the MAOS momenta of neutrinos in the
process $H\rightarrow WW\rightarrow l\nu l^\prime \nu^\prime$. An
interesting feature of  this approach is that one can use the
collider variable $M_{T2}$ \cite{mt2} for event selection, which
enhances both the signal to background ratio and the efficiency of
the MAOS momentum approximation to the true neutrino
momentum\footnote{It has been pointed out recently that $M_{T2}$ can
be used also for the selection of new physics events \cite{BG}.}. As
we will see, the MAOS Higgs mass distribution, constructed with the
MAOS neutrino momenta under a suitable $M_{T2}$ cut, shows a clear
peak over the background at the true Higgs boson mass. One can then
determine the true Higgs boson mass  by performing the likelihood
fit to the MAOS Higgs mass distribution. The results of our
likelihood analysis indicate that the precision can be significantly
improved when the MAOS Higgs momentum is used for the Higgs mass
determination, possibly combined with the kinematic variables
considered in \cite{Aad:2009wy,Barr:2009mx,iwkim}. One can do a
similar likelihood analysis for the Higgs boson discovery or
exclusion, which might enhance the significance of the result for
certain range of the Higgs boson mass.

\section{$M_{T2}$-assisted on-shell (MAOS) momentum}

In this section, we discuss some features of the MAOS  momentum for
the dileptonic Higgs boson decay:
\begin{equation}
H \to \ W(p+k) \, W(q+l) \to \ l(p)\ \nu(k) \ \ l^\prime(q)\
\nu^{\,\prime}(l).
\end{equation}
For the sake of discussion, we decompose the  final state momenta
into the transverse and longitudinal parts as follows:
\begin{eqnarray}
p^\mu &=& (\sqrt{|{\bf p}_T|^2+p_L^2}\,,{\bf p}_T\,,p_L)\,, \ \ \
k^\mu = (\sqrt{|{\bf k}_T|^2+k_L^2}\,,{\bf k}_T\,,k_L)\,, \nonumber \\
q^\mu &=& (\sqrt{|{\bf q}_T|^2+q_L^2}\,,{\bf q}_T\,,q_L)\,, \ \ \ \
l^\mu = (\sqrt{|{\bf l}_T|^2+l_L^2}\ \, \,,{\bf l}_T\,, \ l_L),
\end{eqnarray}
where we have neglected the masses of the charged leptons and
neutrinos. For a Higgs boson mass $m_H \geq 2M_W$, the two $W$
bosons are in on-shell. On the other hand, if $m_H< 2M_W$, one or
both of $W$ bosons should be in off-shell. Regardless of whether the
$W$ bosons are in on-shell or not, one can construct the
event-by-event variable $M_{T2}$ \cite{mt2} which is given
by\footnote{For recent applications of $M_{T2}$ to mass
determination, see \cite{Cho:2007dh,BGL:2007,mt2application}.}
\begin{equation}
M_{T2} \equiv \min_{{\bf k}_T+{\bf l}_T=\psl_T}\left[
\max\left\{M_T^{(1)},M_T^{(2)}\right\} \right]\,,
\end{equation}
where $\psl_T$ denotes the missing transverse momentum carried by
neutrinos, and  $M_T^{(1)\,,(2)}$ are the transverse masses of the
two decaying $W$ bosons, which are given by
\begin{eqnarray}
\label{tmass} \left(M_T^{(1)}\right)^2 = 2\,(|{\bf p}_T||{\bf k}_T|
- {\bf p}_T\cdot{\bf k}_T), \quad \left(M_T^{(2)}\right)^2 =
2\,(|{\bf q}_T||{\bf l}_T| - {\bf q}_T\cdot{\bf l}_T).
\end{eqnarray}

If the Higgs boson is produced without having a sizable transverse
momentum, so that the W-pair transverse momentum ${\bf
p}_T^{WW}\approx 0$, the missing transverse momentum is
(approximately) given by
\begin{equation} \psl_T=-\left({\bf p}_T+{\bf
q}_T\right).\end{equation} In this case,  $M_{T2}$ is simply given
by~\cite{Cho:2007dh}
\begin{equation}
\label{tmaos}M_{T2}^2 = \left(M_T^{(1)}\right)^2 =
\left(M_T^{(2)}\right)^2 = 2\,(|{\bf p}_T||{\bf q}_T| + {\bf
p}_T\cdot{\bf q}_T).
\end{equation}
In fact, for both the signal W bosons from $H\rightarrow WW$ and the
background W bosons from $q\bar{q}\rightarrow WW$, initial state
radiations (ISR) typically give\footnote{Typically the signal has a
slightly larger ${\bf p}_T^{WW}$} $|{\bf p}_T^{WW}|= 10\sim 40$ GeV,
which would be non-negligible compared to the typical lepton
transverse momenta in the final state. Still the above expression of
$M_{T2}$ obtained in the limit ${\bf p}_T^{WW}=0$ shows some
qualitative feature of $M_{T2}$ for non-negligible ${\bf p}_T^{WW}$,
e.g. the correlation between $M_{T2}$ and the transverse opening
angle $\Delta \Phi_{ll}$ between ${\bf p}_T$ and ${\bf q}_T$. In the
following, ISR effects will be fully included in the numerical
evaluation of $M_{T2}$ and the MAOS momentum.

 As the
transverse mass is bounded above by the invariant mass, the above
relation implies
\begin{equation}
M_{T2}\,\leq\, {\rm min}(M^{(1)}, M^{(2)})\, \leq\,
\frac{m_H}{2},\end{equation} where $M^{(1),(2)}$ are the invariant
masses of the intermediate $W$ bosons.  One then immediately finds
that the $M_{T2}$ of dileptonic Higgs decay  is bounded as (upon
ignoring the finite decay widths of the Higgs and W bosons)
\begin{equation}
M_{T2}\,\leq\, {\rm min}(M_W, \frac{m_H}{2}).
\end{equation}
\begin{figure}[t!]
\begin{center}
\epsfig{figure=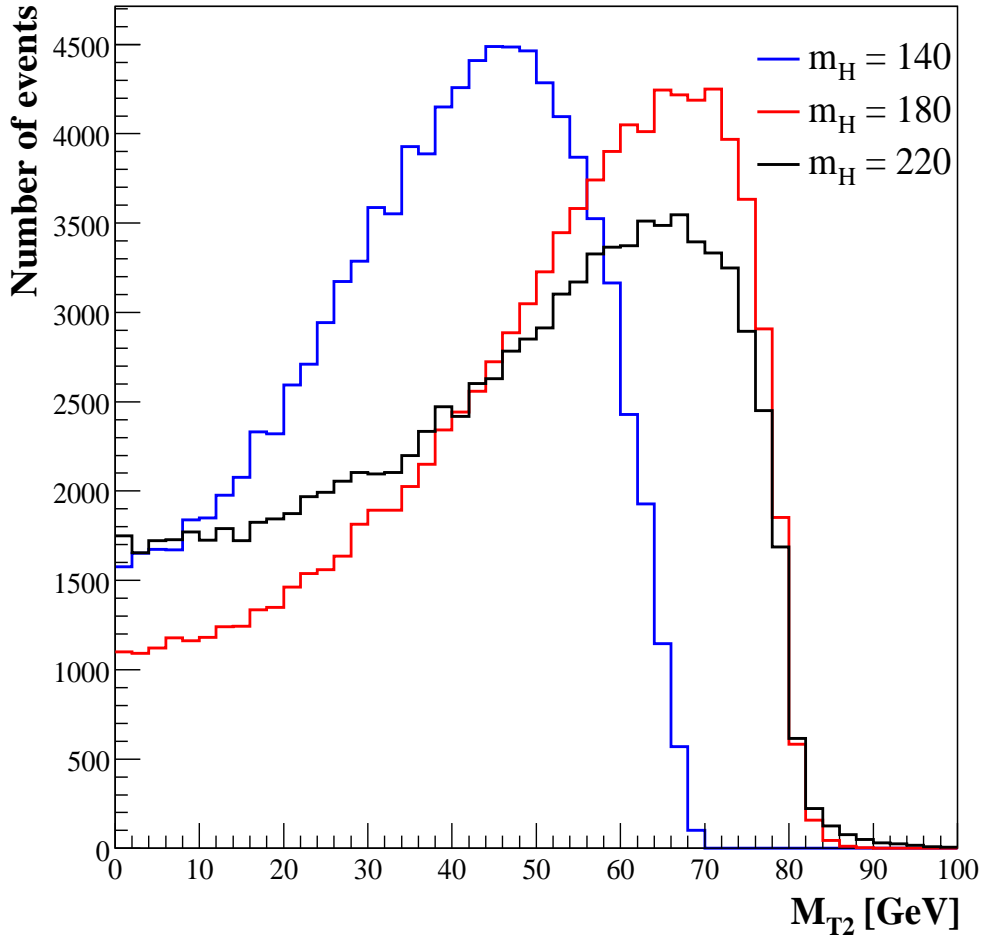,height=14.0cm,width=14.0cm}
\end{center}
\caption{\it The distributions of $M_{T2}$ for various values of $m_H$.}
\label{fig:MT2}
\end{figure}
In Fig. \ref{fig:MT2}, we depict the $M_{T2}$ distribution  for
several different values of $m_H$, where the event set is generated
for the process $gg\rightarrow H\rightarrow WW\rightarrow l\nu
l^\prime \nu^\prime$
 at
the LHC beam condition\footnote{For the analysis in this section, we
do not include the effects of hadronization and detector smearing,
while those effects are incorporated in the analysis of the next
section.}. The results show that the endpoint value is indeed
$m_H/2$ for $m_H< 2M_W$, while
it is $M_W$ for $m_H> 2M_W$.  

The MAOS momenta \cite{Cho:2008tj} are the four-vector variables
which approximate the momenta of invisible particles in collider
event of the type
$$X(p+k)X^\prime(q+l)\rightarrow
V(p)+\chi(k)+V^\prime(q)+\chi^\prime(l),$$ where $X$ and $X^\prime$
denote pair-produced (not necessarily identical and not necessarily
in on-shell) mother particles,  $V(p)$ and $V^\prime(q)$ represent
(one or more) visible particles with total momenta $p$ and $q$,
which are produced by the decays of $X$ and $X^\prime$,
respectively,
 and $\chi$ and $\chi^\prime$ are two invisible particles having the
 same mass.
 The transverse MAOS momenta are defined as the trial transverse
momenta of $\chi$ and $\chi^\prime$ that determine the $M_{T2}$ of
the event, i.e.  ${\bf k}_T$ and ${\bf l}_T$ minimizing
$\max\left\{M_T (X),M_T(X^\prime)\right\}$ under the constraint
${\bf k}_T+{\bf l}_T=\psl_T$, where $M_{T}(X)$ and $M_T(X^\prime)$
are the transverse masses of $X$ and $X^\prime$, respectively.  For
the process $WW\rightarrow l(p)\nu(k) l^\prime(q) \nu^\prime(l)$,
the transverse MAOS momenta are uniquely determined by the
conditions
\begin{eqnarray}
M_{T2}({\bf p}_T,{\bf q}_T,\psl_T)&=&\sqrt{2\,(|{\bf p}_T||{\bf
k}^{\rm maos}_T| - {\bf p}_T\cdot{\bf k}^{\rm maos}_T)}\nonumber
\\
&=&\sqrt{2\,(|{\bf q}_T||{\bf l}^{\rm maos}_T| - {\bf q}_T\cdot{\bf
l}^{\rm maos}_T)}, \nonumber \\
\psl_T&=& {\bf k}^{\rm maos}_T+{\bf l}_T^{\rm maos}.
\end{eqnarray}
For an event without initial state radiation, i.e. an event with
${\bf p}_T^{WW}=0$,  we have $\psl_T=-({\bf p}_T+{\bf q}_T)$ and
$M_{T2}$ given by (\ref{tmaos}). In this case, the transverse MAOS
momenta are simply given by
\begin{equation}
\label{transmaos} {\bf k}^{\rm maos}_T = - {\bf q}_T,\quad {\bf
l}^{\rm maos}_T = - {\bf p}_T.
\end{equation}

There can be two different schemes to define the longitudinal MAOS
momenta. One is to require the on-shell conditions for both the
invisible particles in the final state and the mother particles in
the intermediate state. In the case of $WW\rightarrow l(p)\nu(k)
l^\prime(q) \nu^\prime(l)$, it corresponds to
\begin{equation}
(k_{\rm maos})^2=(l_{\rm maos})^2=0,\quad (p+k_{\rm
maos})^2=(q+l_{\rm maos})^2=M_W^2,\end{equation} which results in
\begin{eqnarray}
\label{longimaos1} {k}^{\rm maos}_L(\pm) &=& \frac{1}{|{\bf
p}_T|^2}\left[ p_L\,A \pm \sqrt{|{\bf
p}_T|^2+p_L^2}\,\sqrt{A^2-|{\bf p}_T|^2|{\bf k}^{\rm maos}_T|^2}
\right]\,, \nonumber \\
{l}^{\rm maos}_L(\pm) &=& \frac{1}{|{\bf q}_T|^2}\left[ q_L\,B \pm
\sqrt{|{\bf q}_T|^2+q_L^2}\,\sqrt{B^2-|{\bf q}_T|^2|{\bf l}^{\rm
maos}_T|^2} \right]\,,
\end{eqnarray}
where $A\equiv M_W^2/2+{\bf p}_T\cdot{\bf k}^{\rm maos}_T$ and
$B\equiv M_W^2/2+{\bf q}_T\cdot{\bf l}^{\rm maos}_T$.
Note that the above longitudinal MAOS momenta have four-fold
degeneracy for each event (two-fold degeneracy  for each neutrino
MAOS momentum).
 Another possible scheme  is to require
\begin{equation}
(k_{\rm maos})^2=(l_{\rm maos})^2=0,\quad (p+k_{\rm
maos})^2=(q+l_{\rm maos})^2=M_{T2}^2,\end{equation}  which gives
unique longitudinal MAOS momenta as
\begin{equation}
\label{longimaos2} k^{\rm maos}_L=\frac{|{\bf k}^{\rm
maos}_T|}{|{\bf p}_T|}p_L,\quad l_L^{\rm maos}=\frac{|{\bf l}^{\rm
maos}_T|}{|{\bf q}_T|}q_L.
\end{equation}
To distinguish these two schemes, the MAOS momenta of
(\ref{transmaos}) and (\ref{longimaos1}) will be called the original
MAOS momenta as they are the one originally defined in
\cite{Cho:2008tj}, while (\ref{transmaos}) and (\ref{longimaos2})
will be called the modified MAOS momenta.

\begin{figure}[t!]
\begin{center}
{\epsfig{figure=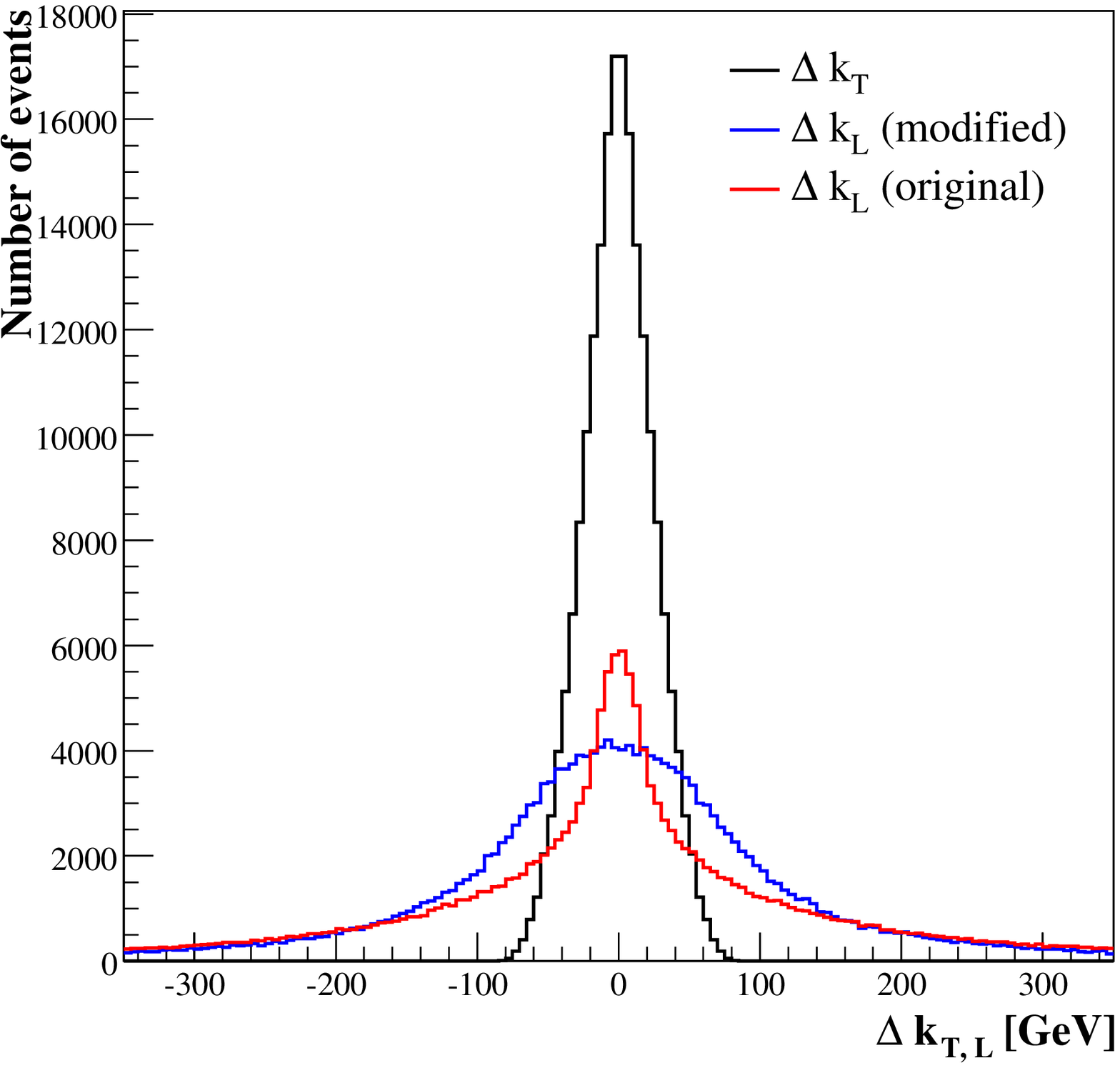,height=7.0cm,width=7.0cm}}
{\epsfig{figure=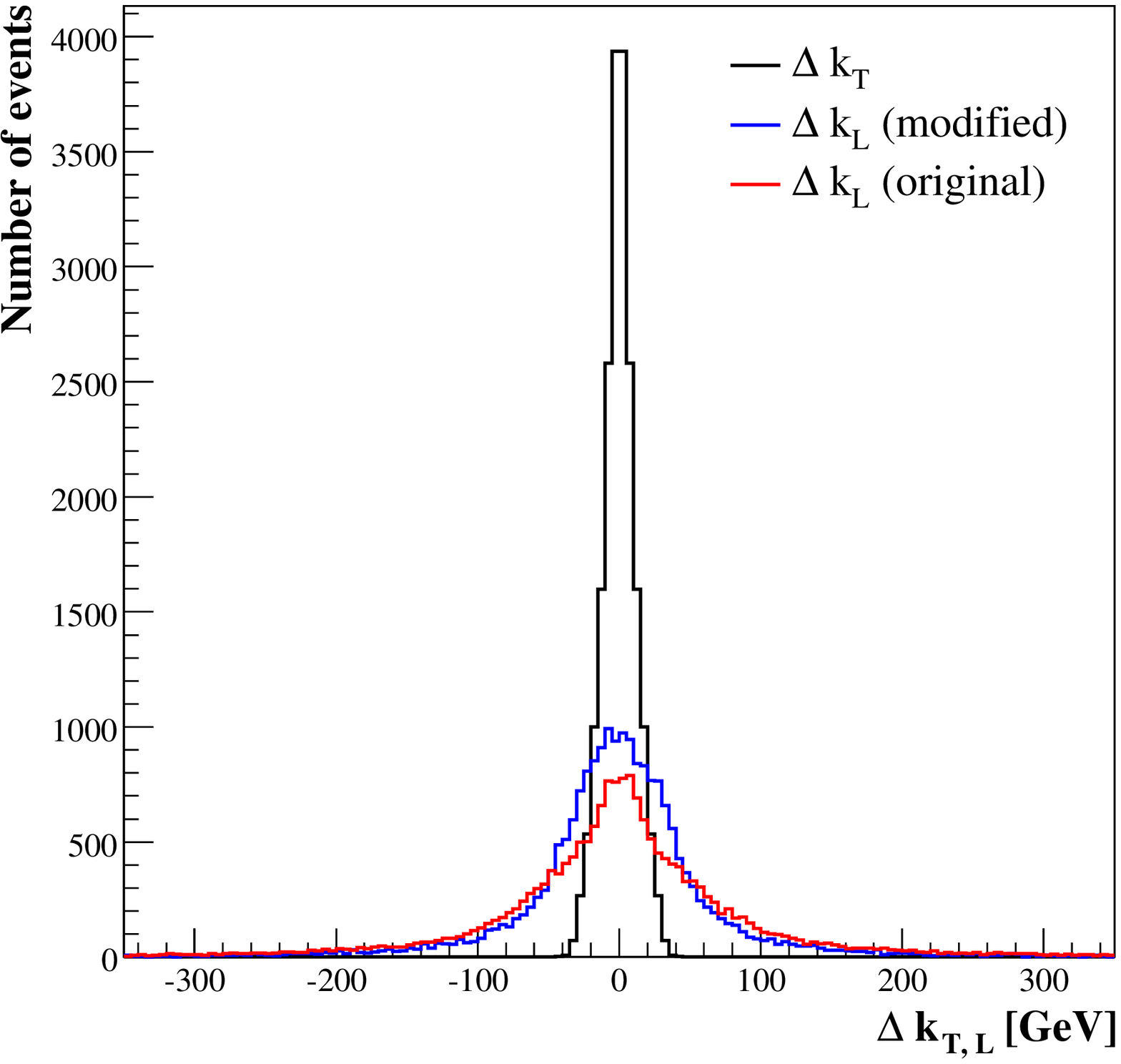,height=7.0cm,width=7.0cm}}
{\epsfig{figure=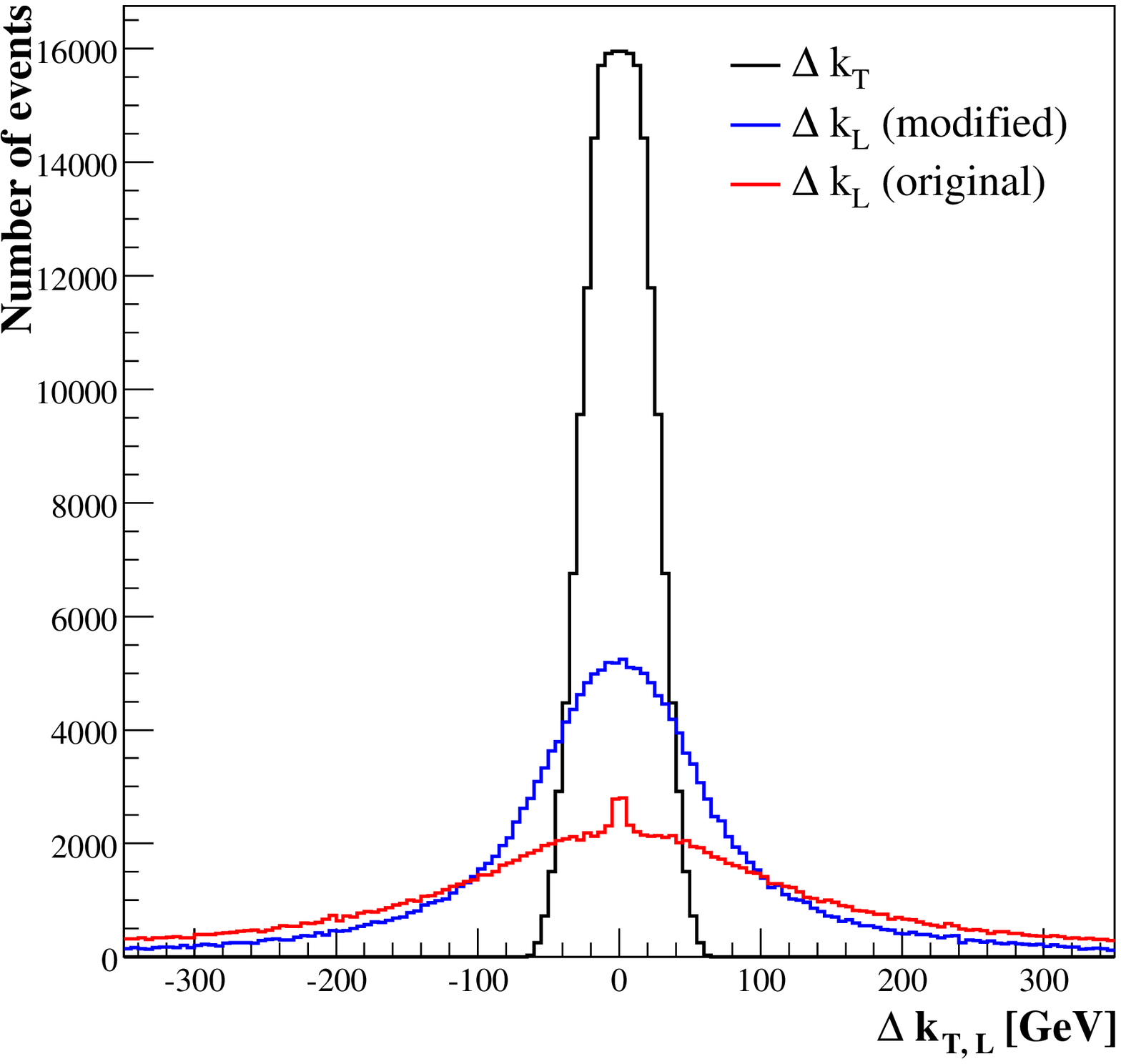,height=7.0cm,width=7.0cm}}
{\epsfig{figure=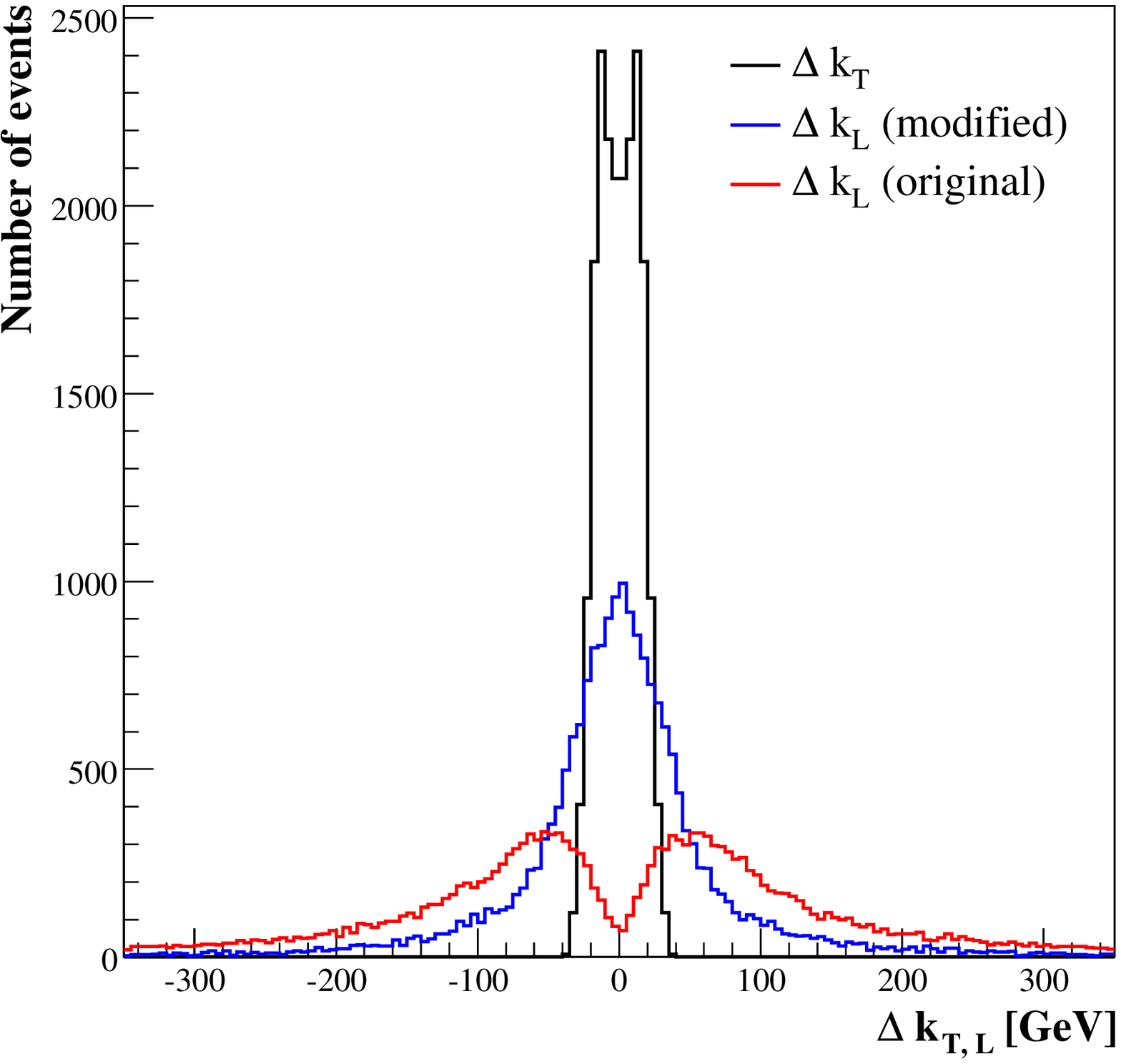,height=7.0cm,width=7.0cm}}
\end{center}
\caption{\it Distributions of $\Delta k_T=k^{\rm true}_T-k^{\rm
maos}_T$ (black), $\Delta k_L$ for  the original MAOS momenta (red),
and $\Delta k_L$ for the modified MAOS momenta (blue).
  The upper frames are for $m_H = 180$ GeV, and the lower
  frames are for $m_H = 140$ GeV. Full event set is used for the left frames, while
 the top $10\,\%$ near endpoint events of $M_{T2}$ are used for the right frames.}
\label{fig:DeltaK}
\end{figure}

A nice feature of the MAOS momenta is that they provide a systematic
approximation to the invisible particle momenta \cite{Cho:2008tj},
i.e. the neutrino momenta in our case. For the case of $m_H>2M_W$,
both of the two $W$ bosons in $H\rightarrow W^+W^-$  are on
mass-shell, and then the endpoint value of $M_{T2}$ is given by
$M_W$. In this case, one easily finds that both the original MAOS
momenta and the modified MAOS momenta approach to the true neutrino
momenta in the limit of the endpoint event with $M_{T2}=M_W$. (Note
that $k_L^{\rm maos}(+)=k_L^{\rm maos}(-)$ for the endpoint event.)
For generic events with $M_{T2}<M_W$, the MAOS momenta generically
differ from the true neutrino momenta. Even in these cases, we can
infer from its distribution that the MAOS momentum provide a
reasonable approximation to the true neutrino momentum.  As the
approximation gets better for larger value of $M_{T2}$, one can
systematically improve the efficiency of approximation with $M_{T2}$
cut.

In Fig. \ref{fig:DeltaK}, we show the distributions of the difference
between the MAOS momentum and the true neutrino momentum,
$$\Delta k_{T,L}=k_{T,L}^{\rm true}-k_{T,L}^{\rm maos},$$ for $m_H=180$
GeV and  $m_H=140$ GeV, respectively. The left panels include the
distributions of the full event set for $gg\rightarrow H\rightarrow
WW\rightarrow l\nu l^\prime \nu^\prime$ generated at the LHC
condition, while the right panels show the distributions of the top
10\% subset near the endpoint of $M_{T2}$. By definition, the
original and modified MAOS schemes give the same transverse MAOS
momenta, thus the same $\Delta k_T$ distribution (black). For
$\Delta k_L$ in the original MAOS scheme, we construct its
distribution using the two solutions $\{k^{\rm maos}_L(+), l^{\rm
maos}_L(+)\}$ and $\{k^{\rm maos}_L(-),l^{\rm maos}_L(-)\}$ for each
event. For $m_H>2M_W$,  our results indicate that the original
scheme (red) is a bit better than the modified scheme (blue), if one
considers the full event set. However, the two schemes show similar
performance if one employs a proper $M_{T2}$ cut, say a cut
selecting about 30\% of the near-endpoint events. Both the modified
and original schemes recover the true neutrino momentum for the
exact endpoint event with $M_{T2}=M_W$. On the other hand, when
$m_H< 2M_W$ so that one or both  $W$-bosons are in off-shell, the
modified MAOS is clearly the better choice to approximate the
neutrino momenta. For off-shell $W$, the original MAOS scheme does
not give correct neutrino momenta even for the endpoint event of
$M_{T2}$, while the modified MAOS scheme does.


\begin{figure}[t!]
\begin{center}
{\epsfig{figure=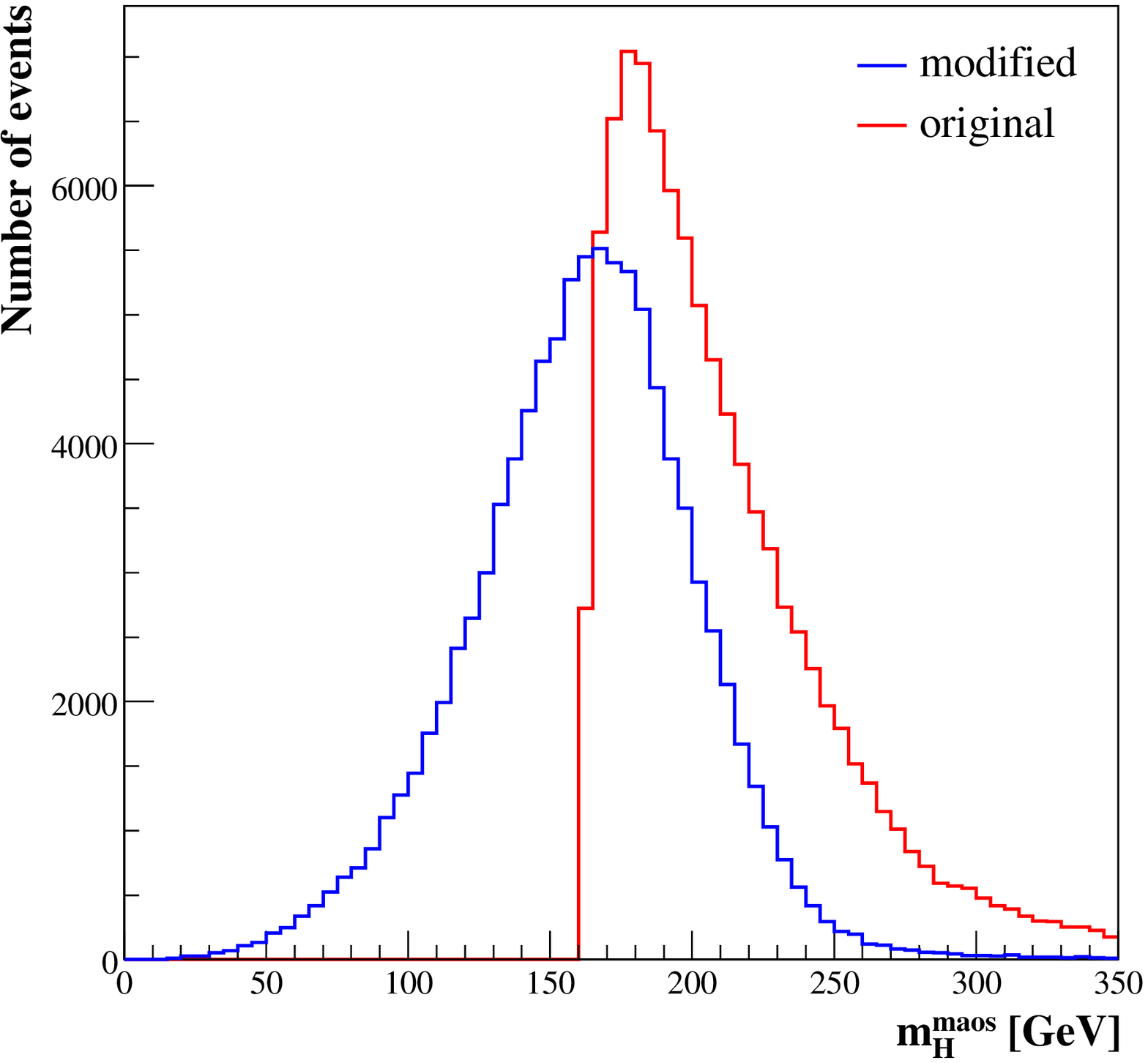,height=7.0cm,width=7.0cm}}
{\epsfig{figure=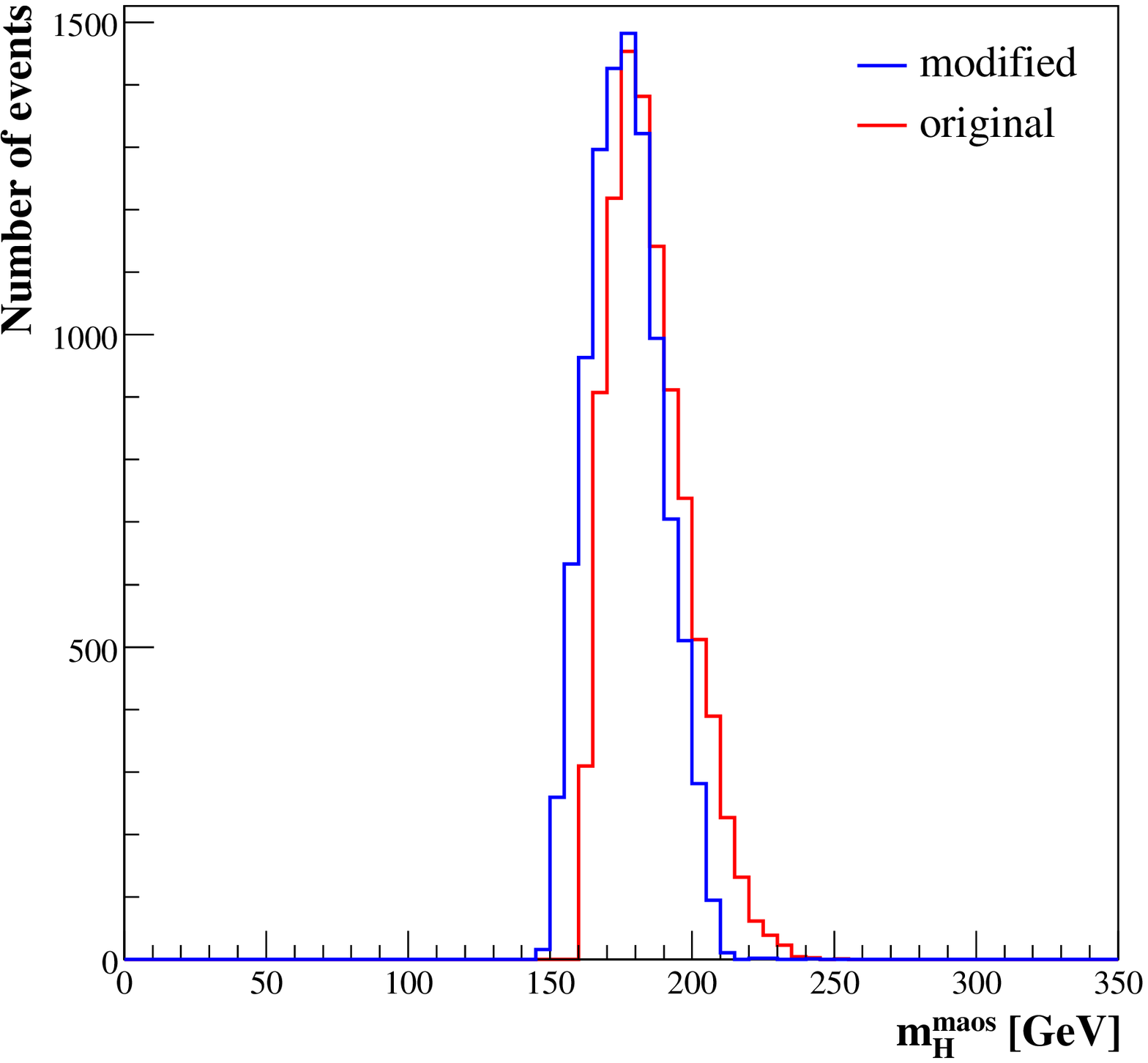,height=7.0cm,width=7.0cm}}
{\epsfig{figure=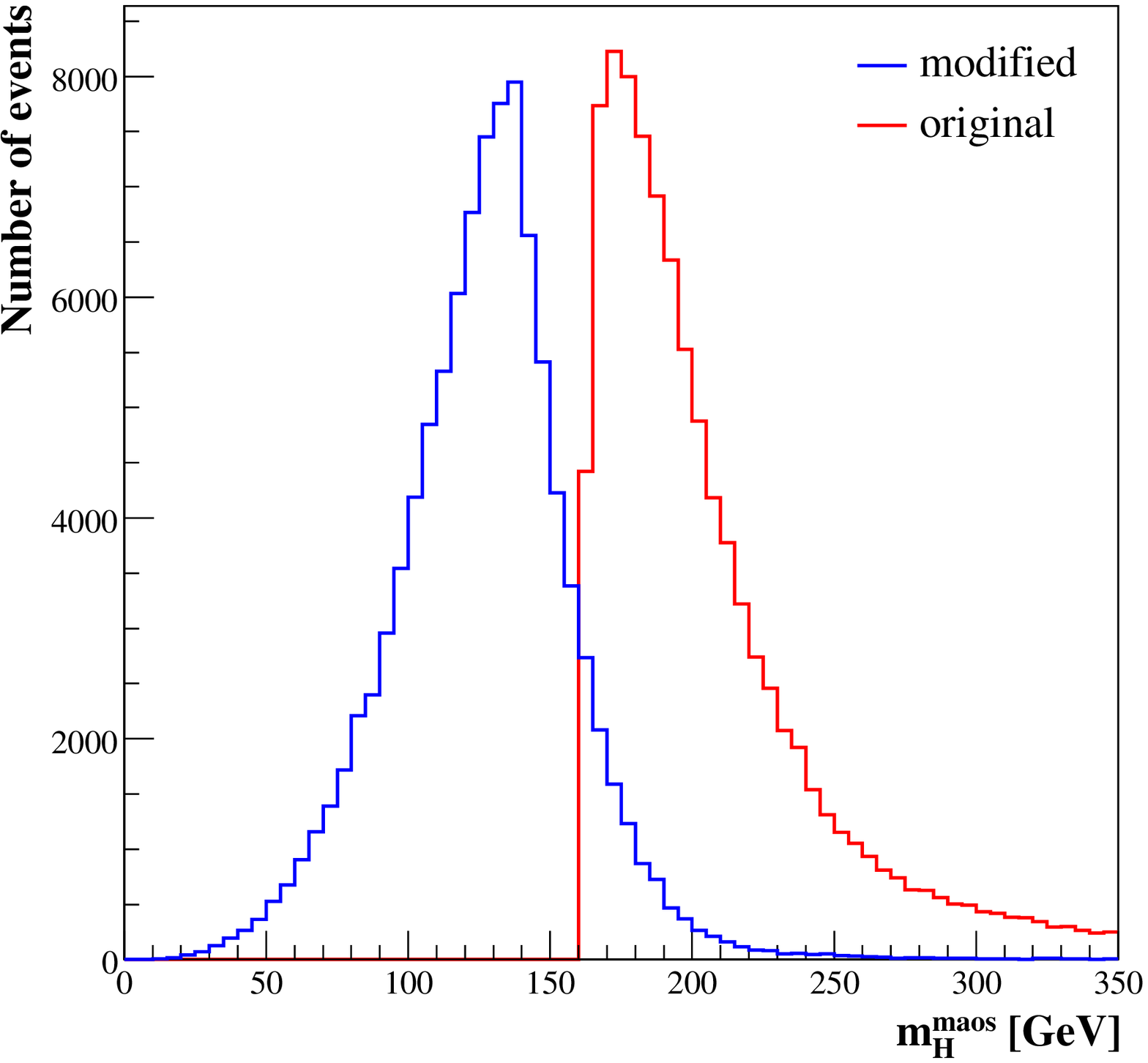,height=7.0cm,width=7.0cm}}
{\epsfig{figure=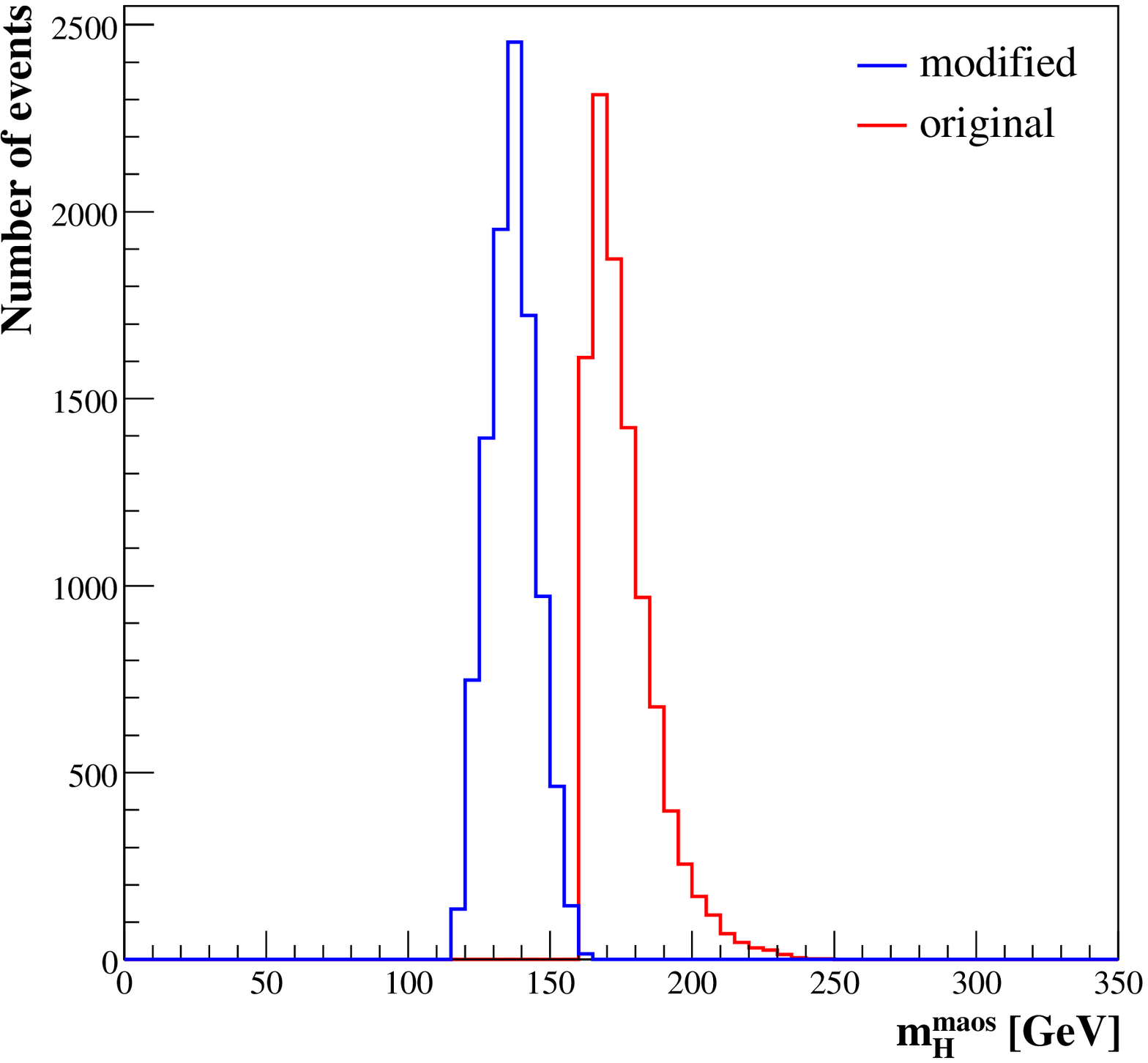,height=7.0cm,width=7.0cm}}
\end{center}
\caption{\it The MAOS Higgs mass distributions for the full event set
  (left frames) and the top 10 \% near endpoint events of $M_{T2}$ (right
  frames)
  in the original (red) and modified (blue) MAOS
  schemes. The upper frames are for $m_H = 180$ GeV  and the
  lower frames are for $m_H = 140$ GeV.}
\label{fig:mhmaos}
\end{figure}

Once the MAOS momenta of neutrinos are obtained, one can construct
the MAOS Higgs mass:
\begin{equation}
\left(m_H^{\rm maos}\right)^2 \equiv (p + {k}_{\rm maos} + q +
{l}_{\rm maos})^2\,.
\end{equation}
A nice feature of $m_H^{\rm maos}$ is that its distribution has a
peak at the true Higgs boson mass, which becomes narrower under a
stronger $M_{T2}$ cut. In Fig. \ref{fig:mhmaos}, we show the MAOS
Higgs mass distributions of the full event set (left panel) of
$gg\rightarrow H\rightarrow WW\rightarrow l\nu l^\prime \nu^\prime$
generated for $m_H=180$ and  $140$ GeV at the LHC condition, and
also of the top 10\% near endpoint events of $M_{T2}$ (right panel),
for both the original MAOS scheme (red) and the modified MAOS scheme
(blue). Our results suggest that the modified MAOS scheme can
provide a good approximation to the invisible neutrino momenta for
both $m_H<2M_W$ and $m_H>2M_W$, under a suitable $M_{T2}$ cut
selecting near endpoint events. In fact, such an $M_{T2}$ cut is
useful in another sense as it enhances the signal to background
ratio. As is well known, the dileptons from the Higgs decay are
likely to have smaller opening angle than the background, which is
essentially due to that the Higgs boson is a spin zero particle.
Then, the expression (\ref{tmaos}) of $M_{T2}$
indicates that the dileptons from the Higgs
decay are likely to have larger $M_{T2}$ than the background because
of the same reason.

In the next section, we will perform the likelihood analysis for the
MAOS Higgs mass distribution to determine the true Higgs boson mass,
while including some of the main backgrounds as well as the detector
effects. We will use the modified MAOS scheme since the original
MAOS scheme does not work for the case of $m_H<2M_W$, while the
modified MAOS scheme works well for both $m_H<2M_W$ and $m_H>2M_W$.

%

\section{MAOS reconstruction of the Higgs boson mass}

To investigate the experimental performance of the MAOS
reconstruction of the Higgs boson mass at the LHC, we use the {\tt
PYTHIA6.4} Monte Carlo (MC) generator at
NLO~\cite{Sjostrand:2006za}.
The generated events have been further processed through the fast
detector simulation program {\tt PGS4}~\cite{pgs4} to incorporate
the detector effects with reasonable efficiencies and fake rates
\cite{conway}.
Assuming the integrated luminosity of 10 fb$^{-1}$, we have
generated the MC event samples of the SM Higgs boson signal and the
two main backgrounds.
For the signal, we consider the Higgs boson production via the gluon
fusion: $gg\rightarrow H$. For the Higgs mass range $130~{\rm GeV}
\lsim m_H \lsim 200$ GeV, the produced Higgs boson decays mainly
into a pair of $W$ bosons.
We take into account  all the dileptonic decay channels of the $W$
bosons to enhance the signal, $W \to l \nu$ with $l=e,\mu$.
The dominant background comes from the continuum $q\bar{q}\,,\,gg
\to WW \to l\nu l^\prime\nu^\prime$ process, and we include also the
$t\bar{t}$ background in which the two top quarks decay into a pair
of $W$ bosons and two $b$ jets.

Following \cite{Aad:2009wy}, we have imposed the following basic
selection cuts on the Higgs signal and the backgrounds:
\begin{itemize}
\item Require that the event has exactly two isolated, opposite-sign leptons (electron or muon) with
$p_T >15$ GeV and $|\eta|<2.5$.
\item 12 GeV $< m_{ll}< 300$ GeV.
\item $|\psl_T| > 30$ GeV.
\item No  $b$ jets.
\item No jets with $p_T > 20$ GeV.
\end{itemize}

\begin{figure}[!t]
\begin{center}
{\epsfig{figure=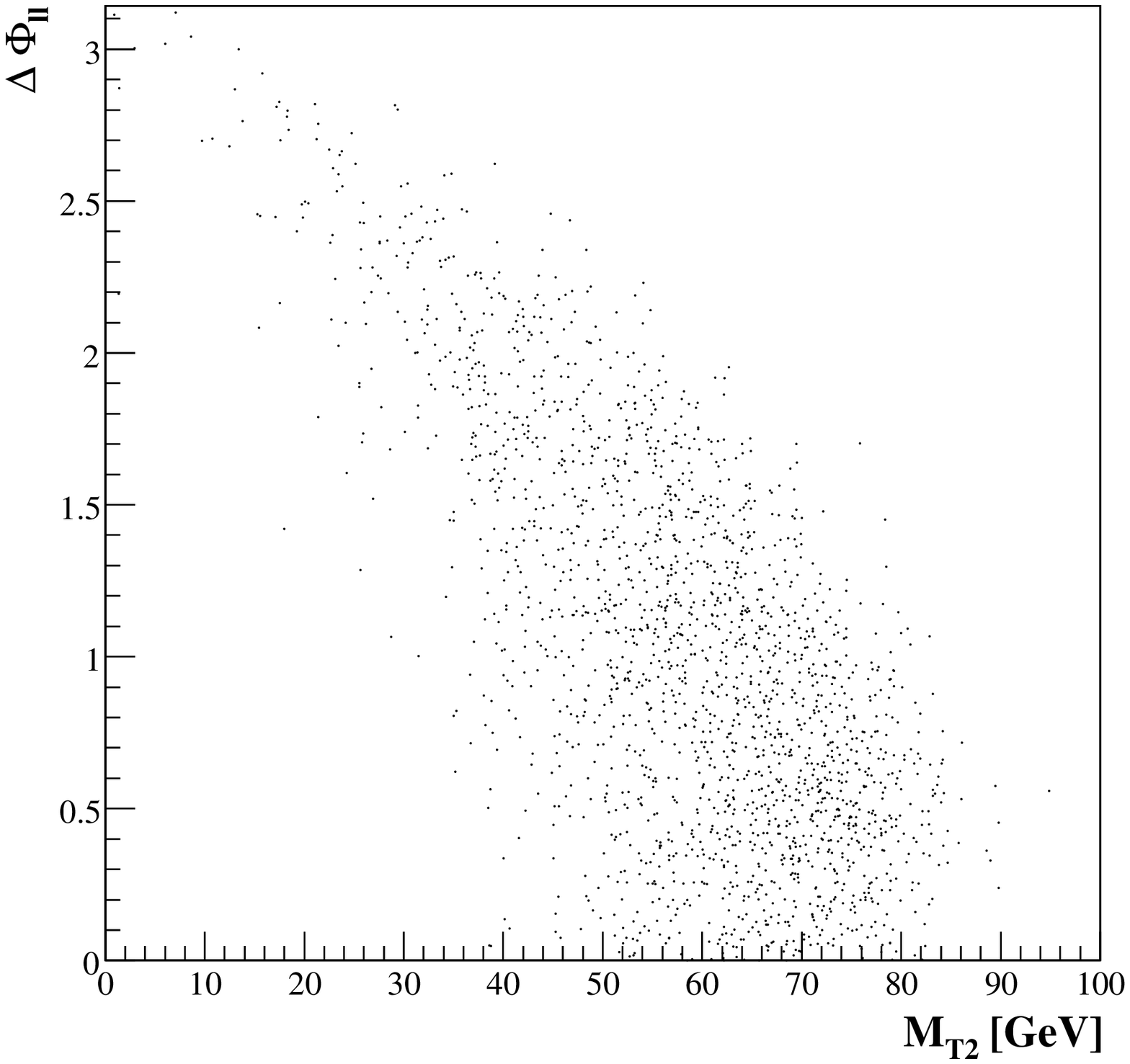,height=7.0cm,width=7.0cm}}
{\epsfig{figure=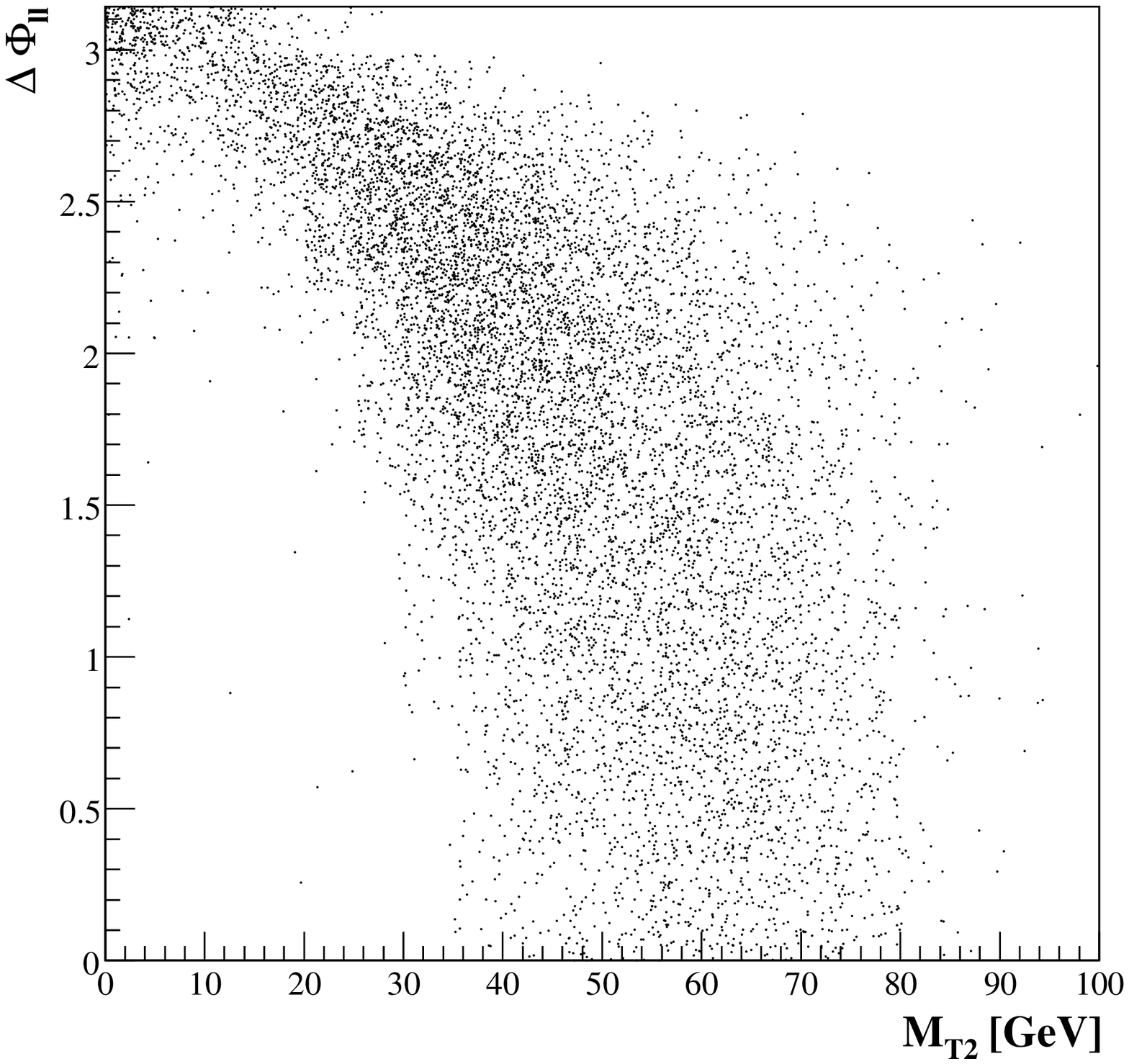,height=7.0cm,width=7.0cm}}
\end{center}
\caption{\it The scatter plots of $M_{T2}$ and $\Delta\Phi_{ll}$ for
the signal (left) and background (right) events after imposing the
basic selection cuts. For the signal, $m_H=170$ GeV is taken. }
\label{fig:DphiMT2_2d}
\end{figure}

\begin{figure}[!t]
\begin{center}
{\epsfig{figure=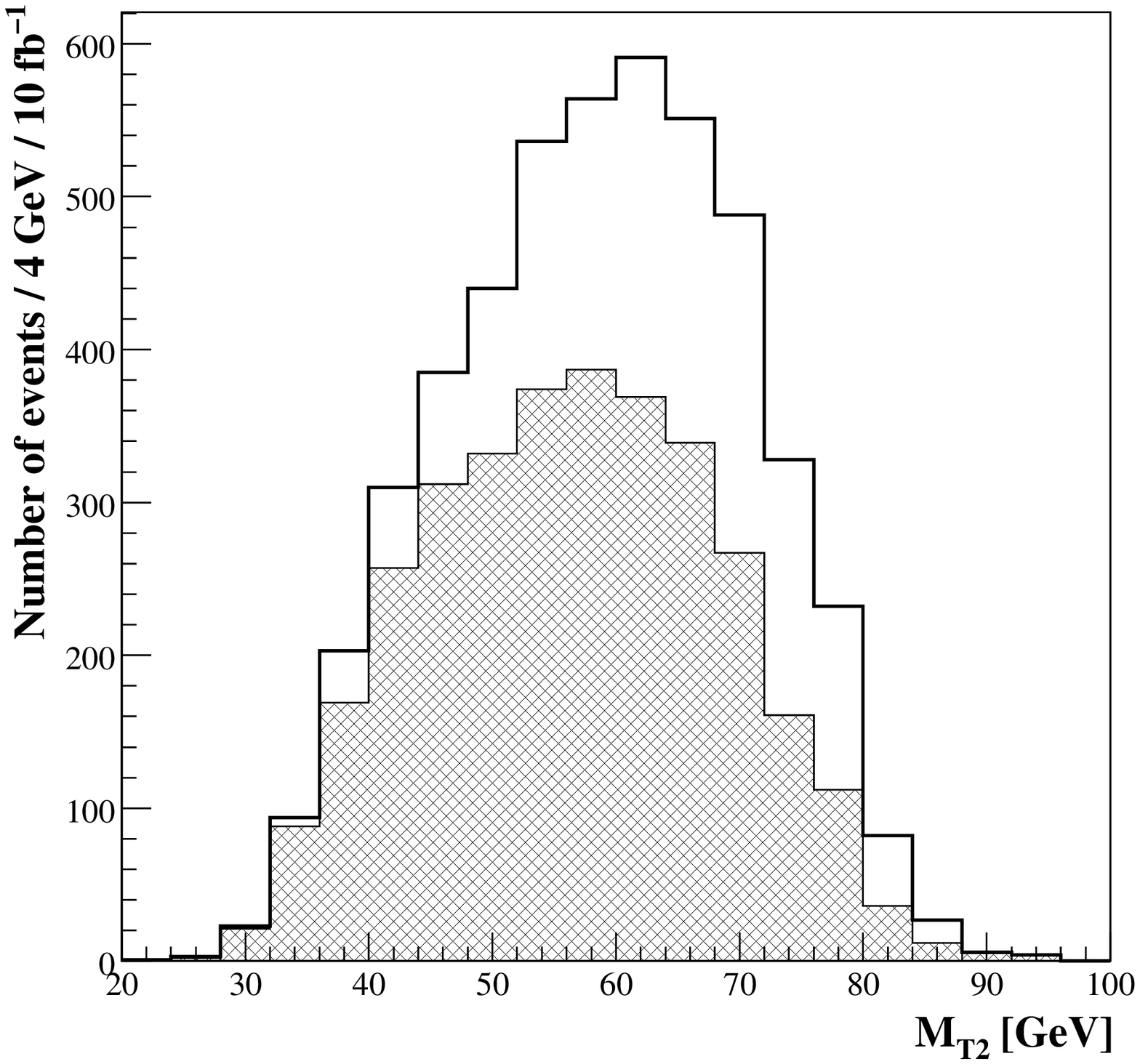,height=7.0cm,width=7.0cm}}
{\epsfig{figure=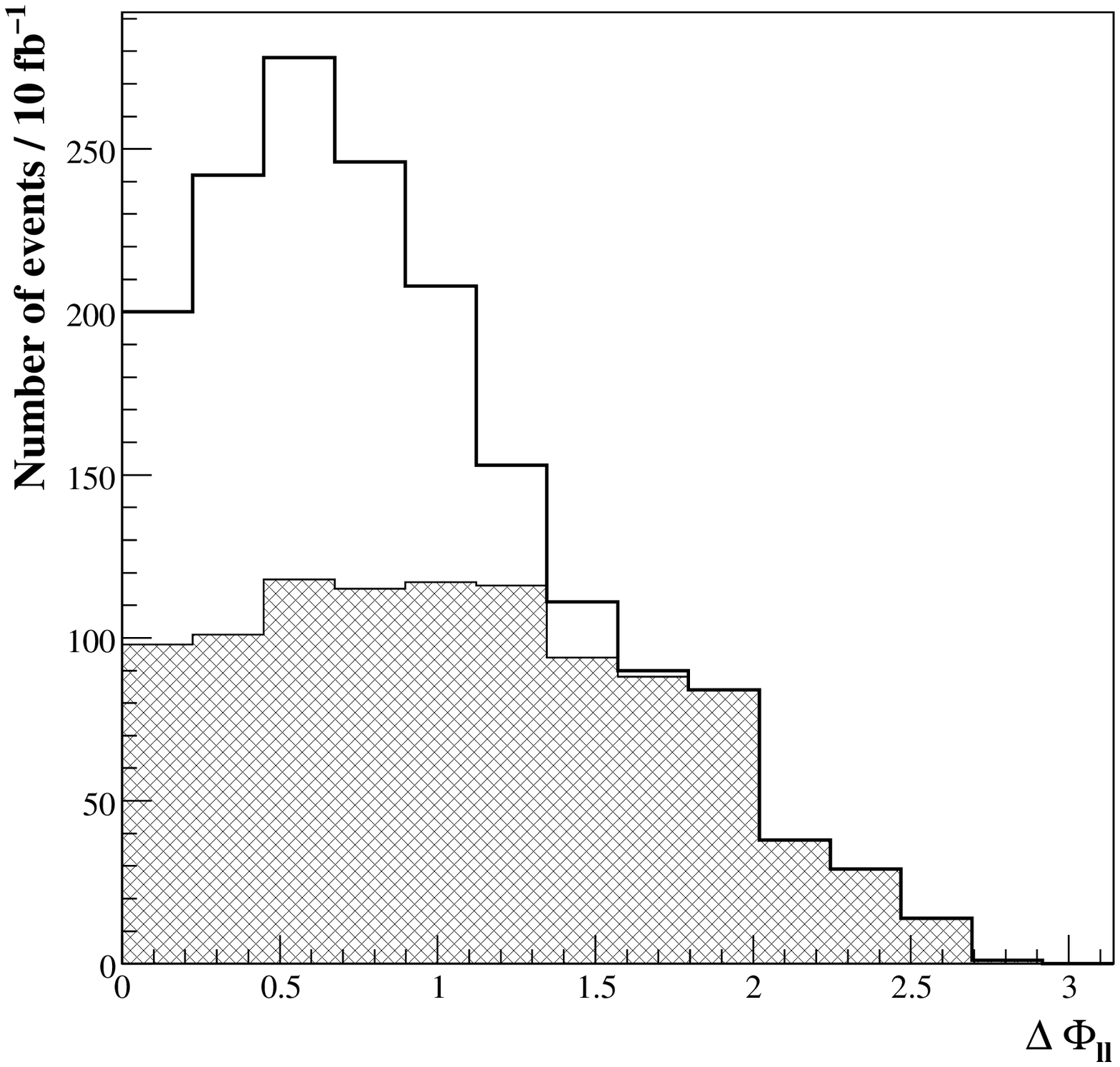,height=7.0cm,width=7.0cm}}
\end{center}
\caption{\it The $M_{T2}$ distribution (left) for the events with
$\Delta\Phi_{ll} < 1.6$,  and the $\Delta\Phi_{ll}$ distribution
(right) for the events with $M_{T2} > 67$ GeV. Signal events are
taken for $m_H=170$ GeV,  and the shaded regions represent  the
backgrounds.} \label{fig:DphiMT2}
\end{figure}

It is well known that the background can be significantly reduced by
exploiting the helicity correlation between the charged lepton and
its mother $W$ boson. Introducing the transverse opening angle
between  two charged leptons, $\Delta\Phi_{ll}$, the Higgs signal
tends to have a smaller $\Delta\Phi_{ll}$ than the background, which
is essentially due to the fact that the Higgs boson is a spin zero
particle.
%
Selecting the events with large value of $M_{T2}$ similarly enhances
the signal to background ratio, which can be understood by the
correlation between $M_{T2}$ and $\Delta \Phi_{ll}$ in
(\ref{tmaos}). In our case, this $M_{T2}$ cut is particularly useful
since it enhances also the accuracy of the MAOS reconstruction of
the neutrino momenta as discussed in the previous section.
In Fig.~\ref{fig:DphiMT2_2d}, we show the scatter plots of $M_{T2}$
and $\Delta\Phi_{ll}$ for the signal (left panel) and the background
(right panel), obtained after imposing the above basic selection
cuts to the data set for $m_H=170$ GeV, while including the detector
effects. As anticipated, the low $\Delta\Phi_{ll}$ and high $M_{T2}$
region is more populated by the signal events, while the high
$\Delta\Phi_{ll}$ and low $M_{T2}$ region by the backgrounds. We can
also notice a correlation between $\Delta\Phi_{ll}$ and $M_{T2}$
suggested by (\ref{tmaos}). Fig.~\ref{fig:DphiMT2} shows the
$M_{T2}$ distribution (left panel), again for $m_H=170$ GeV, of the
events with $\Delta\Phi_{ll}\leq 1.6$, and the $\Delta\Phi_{ll}$
distribution (right panel) of the events with $M_{T2}\geq 67$ GeV,
where the shaded regions represent the backgrounds. Here, the tail
of the $M_{T2}$ distribution beyond $M_W$ is mainly due to the
$W$-boson width. We observe that the signal is more likely to have
larger $M_{T2}$. Furthermore, the right panel of
Fig.~\ref{fig:DphiMT2} indicates that the $M_{T2}$ cut can
significantly enhance the efficiency of the $\Delta \Phi_{ll}$ cut.
We thus introduce two additional cuts:
\begin{itemize}
\item $\Delta\Phi_{ll} < \Delta\Phi_{ll}^{\rm cut},$
\item $M_{T2} > M_{T2}^{\rm cut},$
\end{itemize}
where  $\Delta \Phi_{ll}^{\rm cut}$ and $M_{T2}^{\rm cut}$ are
chosen to optimize the Higgs mass measurement using the MAOS mass
distribution, and their values for various $m_H$ are listed in
 Table~\ref{tab:tuned_cuts}.

%


%

\begin{table}[\hbt]
\caption{\label{tab:tuned_cuts} {\it The $\Delta\Phi_{ll}$ and
$M_{T2}$ cuts for various values of $m_H$. }}
\begin{center}
\begin{tabular}{ c|rrrrrrrr }\hline\hline
&&&&\\[-2mm]
$m_H$ (GeV) & 130~ & 140~ & 150~ & 160~ & 170~ & 180~ & 190~ & 200~  \\[2mm]
\hline
&&&&\\[-2mm]
$\Delta\Phi_{ll}^{\rm cut}$ & 1.85 & 1.70 & 1.65 & 1.50 & 1.60 & 1.70 & 1.90 & 2.05  \\[2mm]
$M_{T2}^{\rm cut}$ (GeV)
 & 38.0 & 51.0 & 57.0 & 66.0 & 67.0 & 68.0 & 69.5 & 70.0  \\[2mm]
\hline\hline
\end{tabular}
\end{center}
\end{table}
\begin{table}[\hbt]
\caption{\label{tab:cutflow} {\it Cut flows for $m_H=170$ GeV  with
$\Delta\Phi_{ll}^{\rm cut}=1.6$ and $M_{T2}^{\rm cut}=67$ GeV at 10
$fb^{-1}$.
}}
\begin{center}
\begin{tabular}{ l|c|rrr }\hline\hline
&&&&\\[-2mm]
Selection & Selection cuts & $gg \to H$  & $WW~$  & $t\bar{t}~~~~$  \\[2mm]
\hline
&&&&\\[-2mm]
&Lepton selection $+ m_{ll}$ &  $4,445$ & $18,501$ & $139,256$ \\[2mm]
Basic&$|\psl_T|>30$ GeV &  $4,012$ & $12,801$ & $120,597$ \\[2mm]
Selection&$b$-veto  &  $3,956$ & $12,656$ & $60,438$ \\[2mm]
&Jet veto  &  $2,039$ & $8,096$ & $1,287$ \\[2mm]
\hline
&&&&\\[-2mm]
Tuned&$\Delta\Phi_{ll} < \Delta\Phi_{ll}^{\rm cut}$ &  $1,621$ & $2,939$ & $332$ \\[2mm]
Selection&$M_{T2} > M_{T2}^{\rm cut}$ &  $619$ & $585$ & $107$ \\[2mm]
\hline\hline
\end{tabular}
\end{center}
\end{table}
In Table~\ref{tab:cutflow}, taking the case of $m_H=170$ GeV as a
specific example, we show how the numbers of signal events and
background events are changing under each selection cut.
Comparing with the ATLAS cut flows reported in \cite{Aad:2009wy},
the signal and the dominant $WW$ background are in excellent
agreement except for that the overall number of events  is somewhat
larger in our case\footnote{Note that we consider all of the
$ee,\mu\mu,e\mu$ events, while \cite{Aad:2009wy} included only the
$e\mu$ events.}. This may be attributed to the differences in
simulating the jet reconstruction, detector effects and triggering.
For the $t\bar{t}$ background, we have a sizable number of events
even after imposing the basic selection cuts,  but it can be
suppressed to a negligible level by the $\Delta\Phi_{ll}$ and
$M_{T2}$ cuts.
%

\begin{figure}[!t]
\begin{center}
{\epsfig{figure=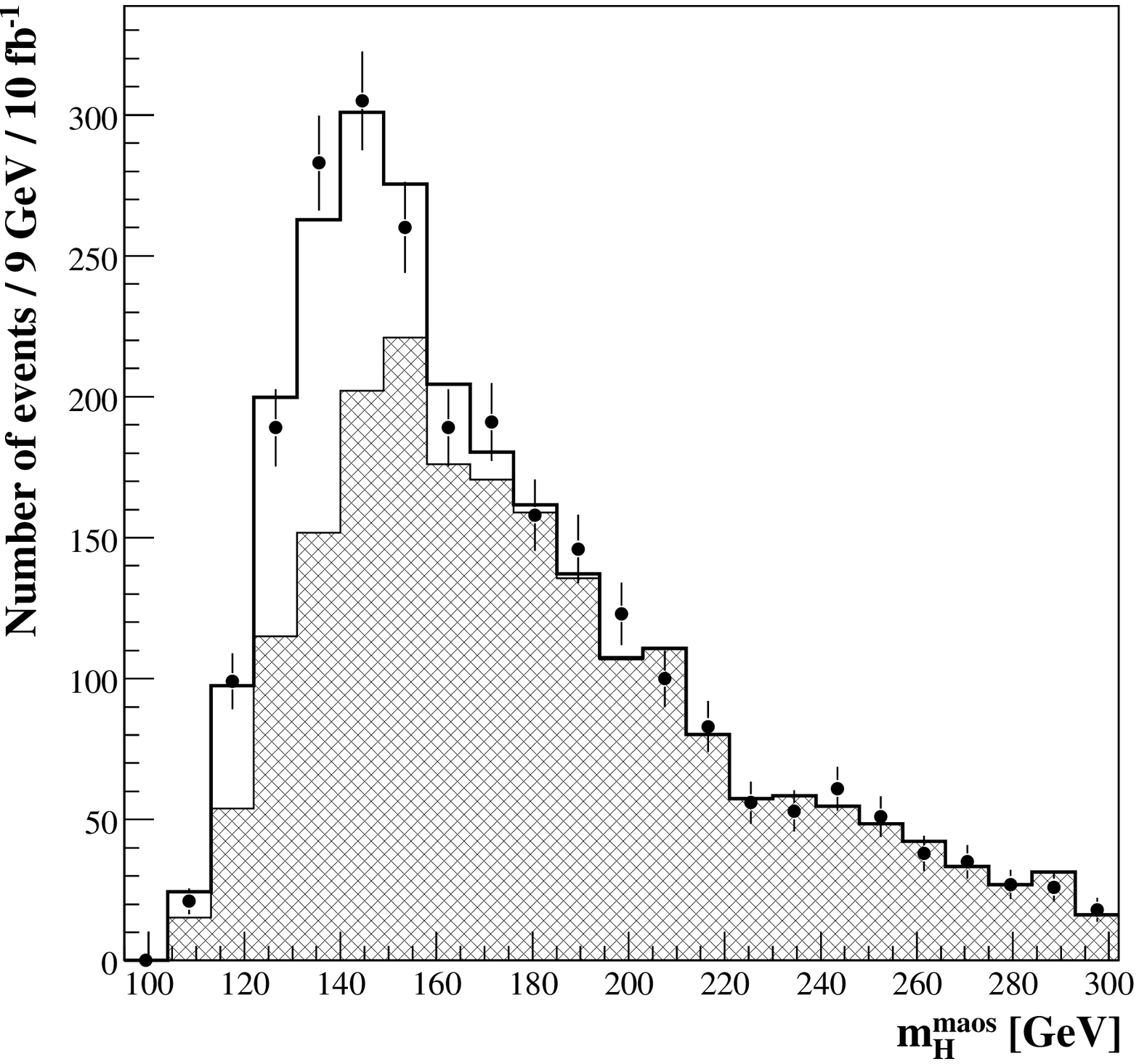,height=6.7cm,width=6.7cm}}
{\epsfig{figure=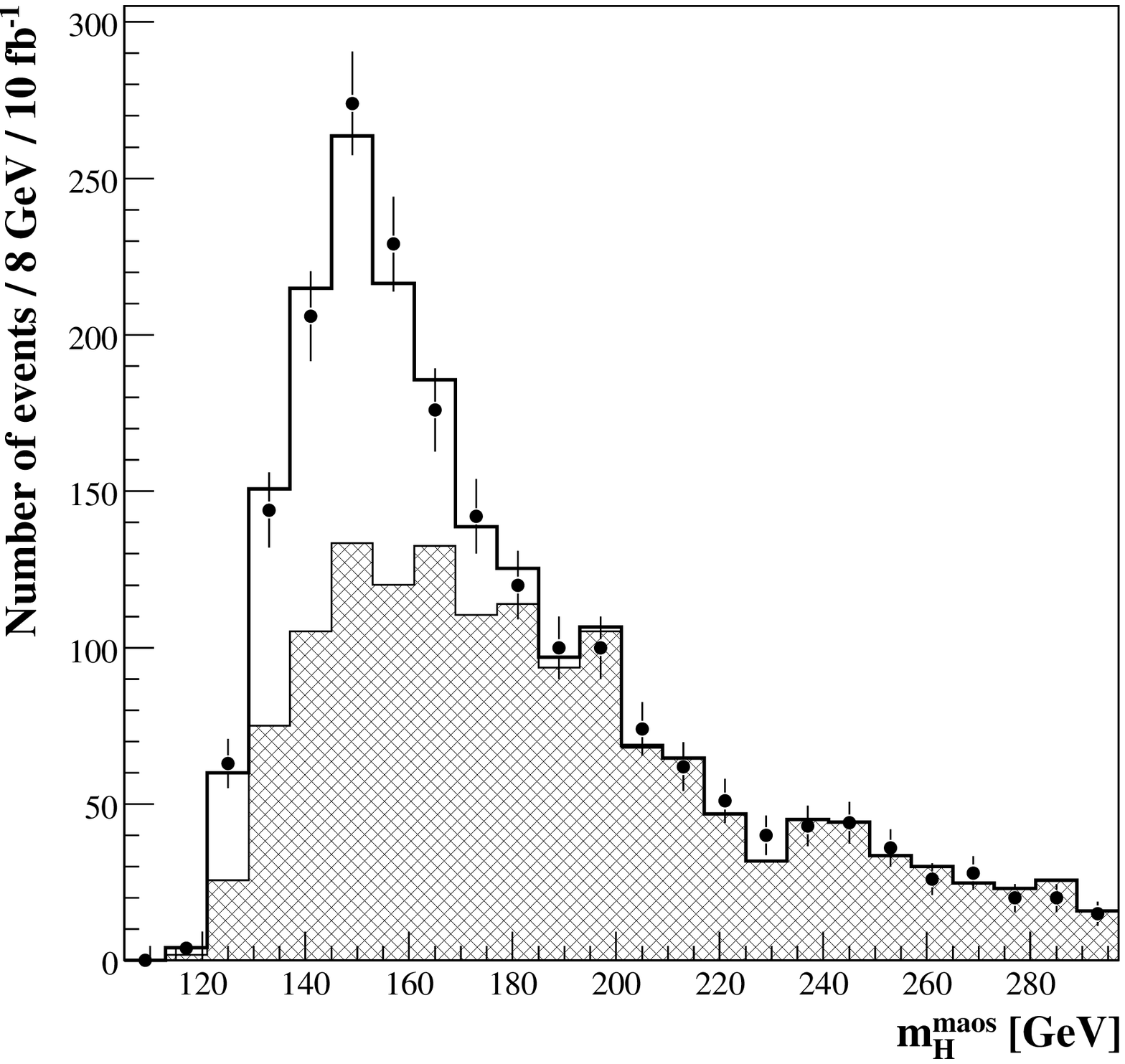,height=6.7cm,width=6.7cm}}
{\epsfig{figure=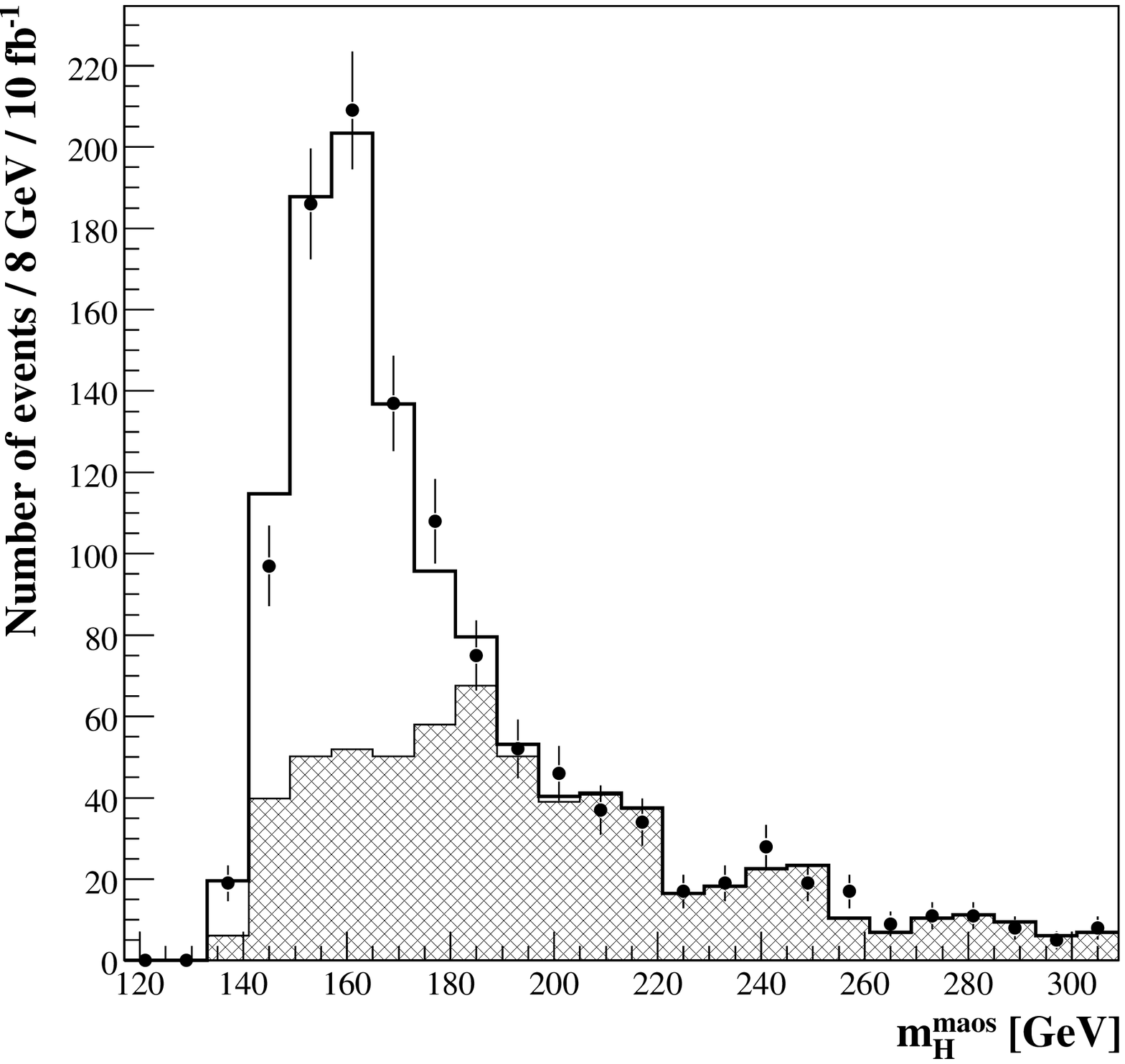,height=6.7cm,width=6.7cm}}
{\epsfig{figure=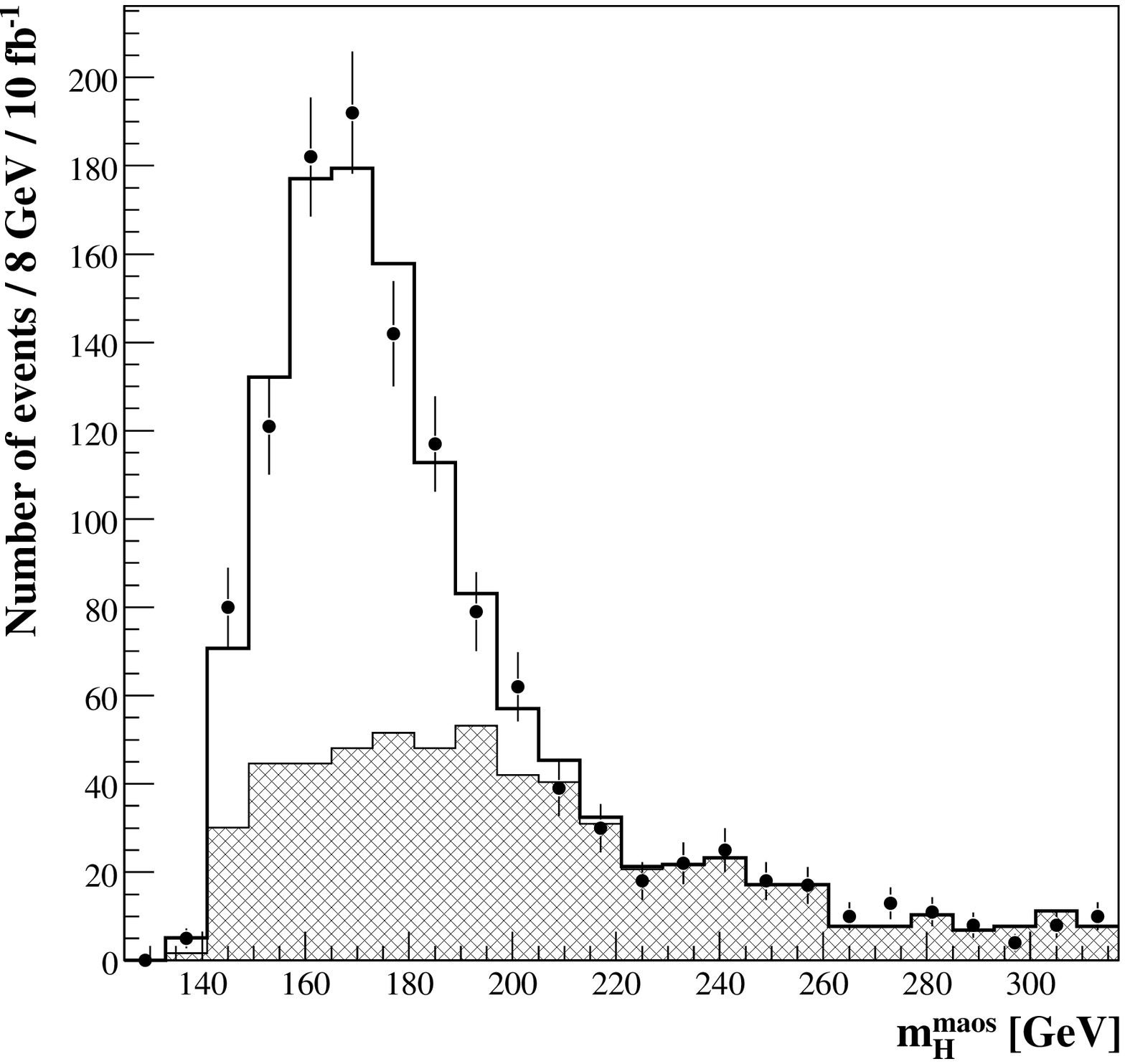,height=6.7cm,width=6.7cm}}
{\epsfig{figure=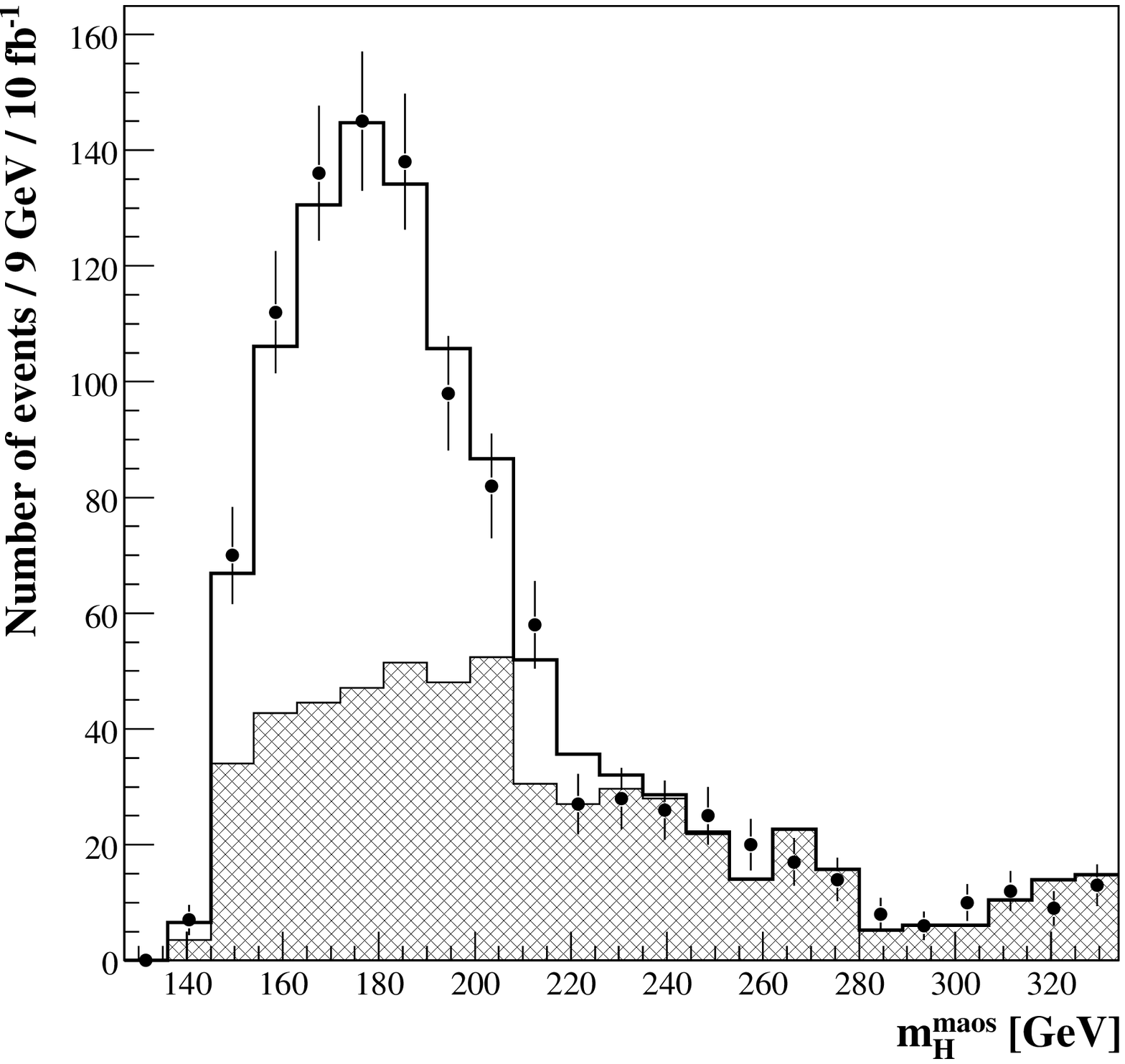,height=6.7cm,width=6.7cm}}
{\epsfig{figure=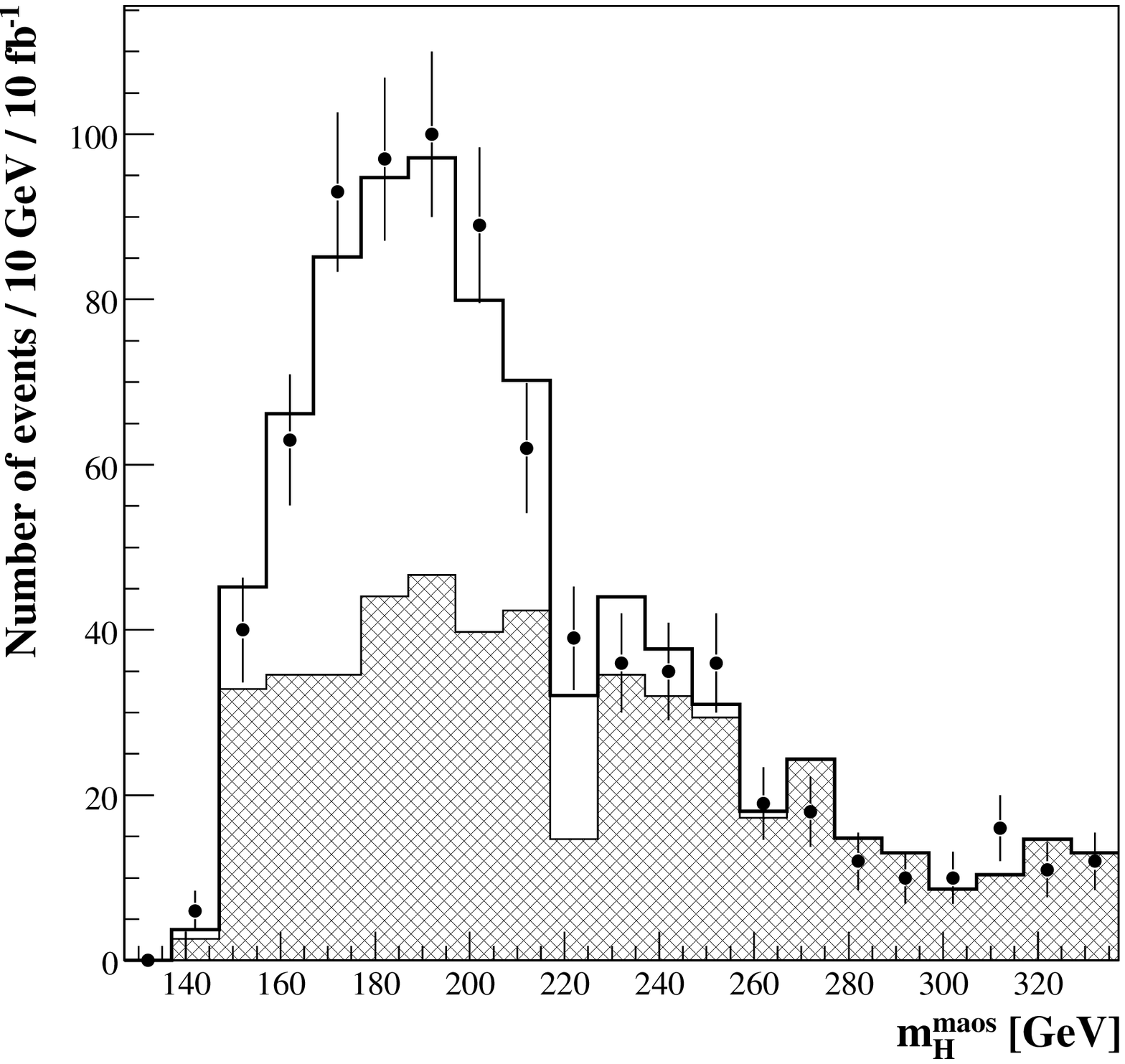,height=6.7cm,width=6.7cm}}
\end{center}
\caption{\it The $m_H^{\rm maos}$ distribution for the nominal data
(dots) and the template (solid line) for various Higgs boson masses:
$m_H=140, \,150, \,160,\, 170,\, 180,\, 190$ GeV from the top left
to the bottom right. In each frame, shaded region represents the
backgrounds.} \label{fig:mh}
\end{figure}

In Fig.~\ref{fig:mh}, we show the ${m}_H^{\rm maos}$ distributions
for various (nominal) values of the Higgs boson mass, obtained by
the modified MAOS momenta of neutrinos, while incorporating the
$M_{T2}$ and $\Delta \Phi_{ll}$ cuts listed in
Table~\ref{tab:tuned_cuts}.
 We observe that the
modified MAOS scheme is working nicely and each distribution has a
clear peak at the true (nominal) Higgs mass independently of whether
$m_H$ is below or above $2 M_W$.

With the ${m}_H^{\rm maos}$ distribution constructed as above, we
performed a template fitting to determine the Higgs boson mass.
Here a template means a simulated distribution with a trial Higgs
mass which, in general, is different from the nominal one used to
generate the data. For each distribution with the nominal Higgs mass
$m_H$, the 11 templates are generated with the trial Higgs masses
between $m_H-10$ GeV and $m_H+10$ GeV, in steps of 2 GeV. For
example, in each frame of Fig.~\ref{fig:mh}, the solid line shows
the template when the trial Higgs mass is the same as the nominal
one. Each template is normalized to the corresponding nominal
distribution.

\begin{figure}[!t]
\begin{center}
{\epsfig{figure=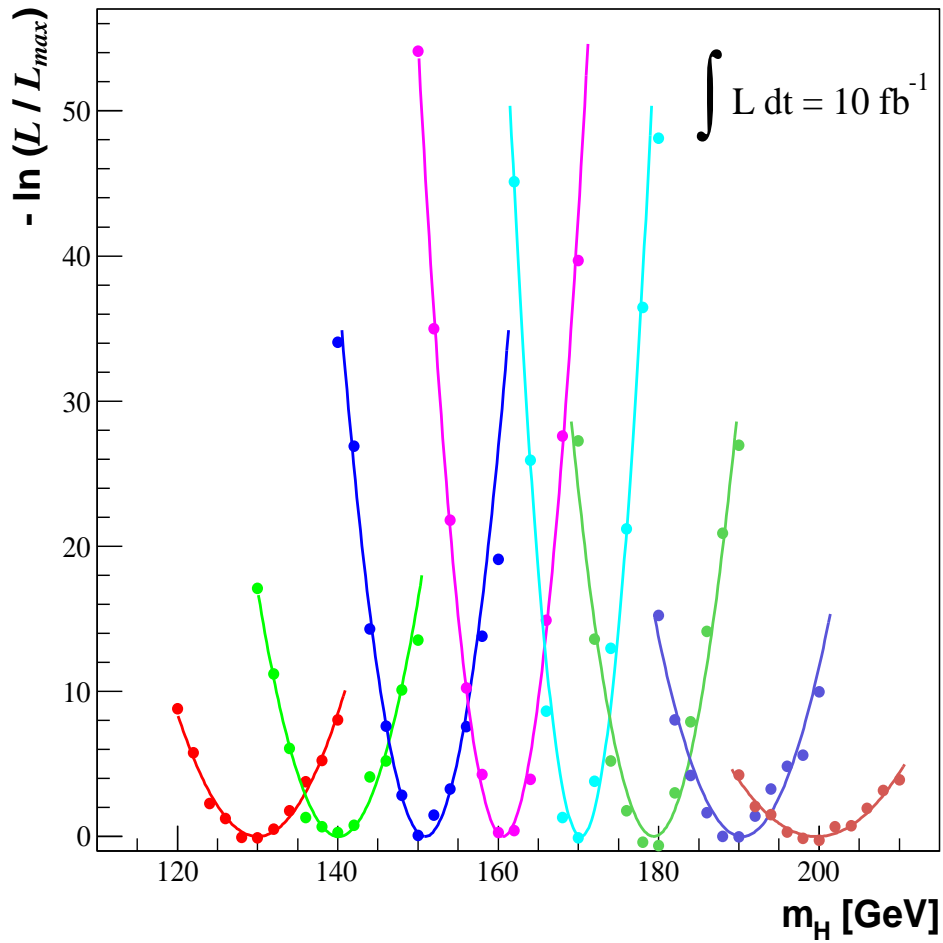,height=14.0cm,width=14.0cm}}
\end{center}
\caption{\it The relative log likelihood distributions for various
  Higgs boson masses. The solid line shows the result of the quadratic
  fitting for each value of $m_H$. The $\mathcal{L}_{\rm max}$ is the
  maximum likelihood which was determined as the minimum of a fit to
  the $-\ln\mathcal{L}$ distribution.} \label{fig:likelihood}
\end{figure}

\begin{figure}[!t]
\begin{center}
{\epsfig{figure=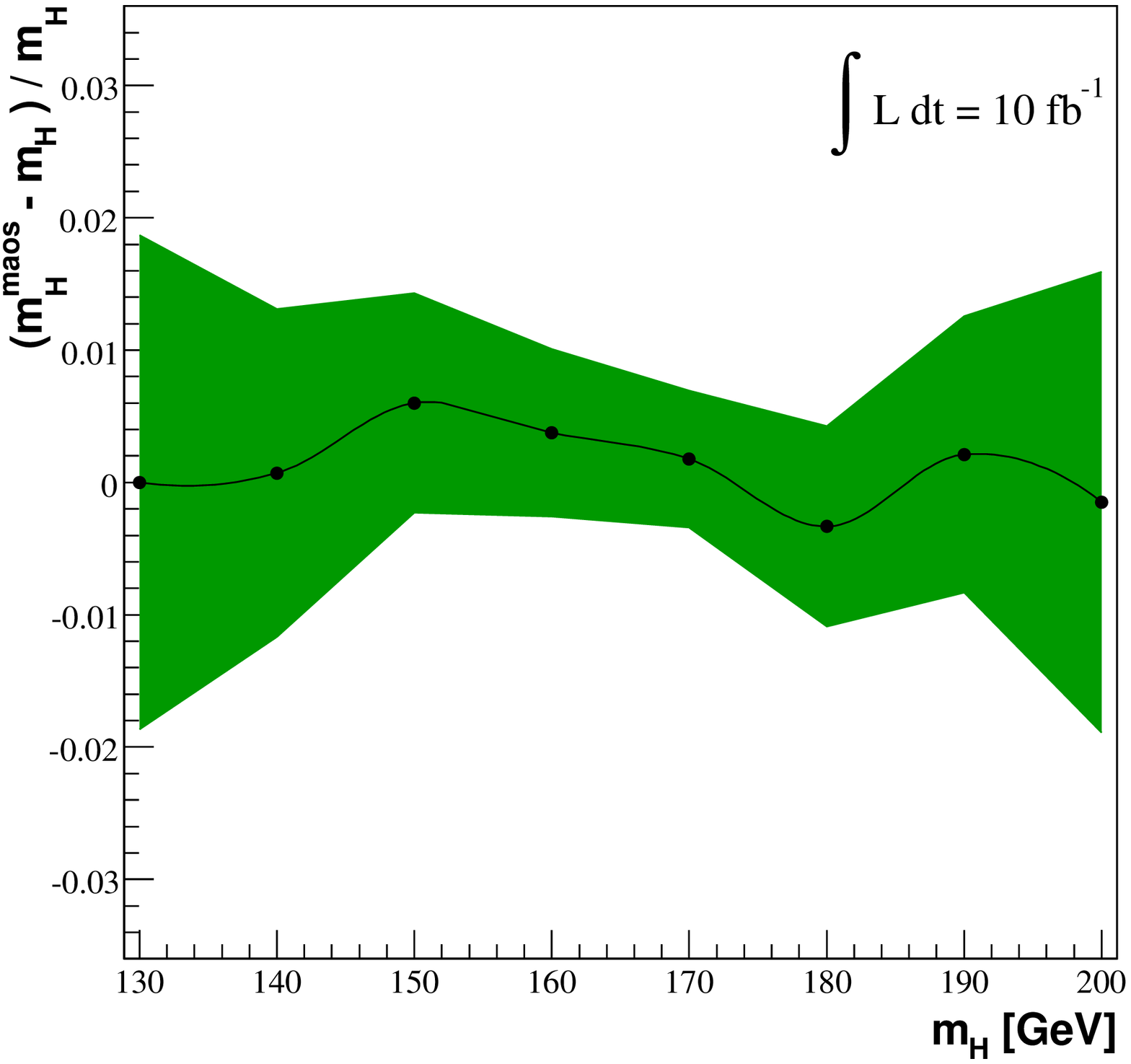,height=14.0cm,width=14.0cm}}
\end{center}
\caption{\it The band showing the 1-$\sigma$ deviation error for the
Higgs boson mass determined by the $m_H^{\rm maos}$ distribution.
The dots and lines denote the Higgs boson mass obtained by the
likelihood fit.  } \label{fig:deviation}
\end{figure}

The likelihood between a nominal data distribution and a template
is defined as the product of individual Poisson probabilities
computed in each bin $i$ over the ${\cal N}$ bins in the fit range:
\begin{equation}
\mathcal{L}\equiv \prod^{\cal N}_{i} \frac{{\rm e}^{-m_i}\, m_i^{n_i}}{n_i !}\,,
\end{equation}
where $n_i$ and $m_i$ denote the number of events in the $i$-th bin
of the nominal distribution and the normalized template,
respectively. In Fig.~\ref{fig:likelihood}, we show the log
likelihood distribution for various Higgs masses in the range
between 130 GeV and 200 GeV. The solid line shows the result of a
quadratic fitting for each value of $m_H$. The fitted
Higgs boson masses together with 1-$\sigma$ error are listed in
Table~\ref{tab:mh_fitted} (see also Fig.~\ref{fig:deviation})
 for various input Higgs masses,
where the 1-$\sigma$ deviated value is defined as the one increasing
$-\ln\mathcal{L}$ by $1/2$~\cite{Amsler:2008zzb}.
%
%

\begin{table}[\hbt]
\caption{\label{tab:mh_fitted} {\it The fitted Higgs boson mass for
various input values of $m_H$ at 10 fb$^{-1}$. }}
\begin{center}
\begin{tabular}{ c|rrrrrrrr }\hline\hline
&&&&\\[-2mm]
$M_H$ (GeV) & 130~ & 140~ & 150~ & 160~ & 170~ & 180~ & 190~ & 200~  \\[2mm]
\hline
&&&&\\[-2mm]
Fitted value (GeV) & 130.0 & 140.1 & 150.9 & 160.6 & 170.3 & 179.4 & 190.4 & 199.7  \\[2mm]
1-$\sigma$ error (GeV)
 & 2.4 & 1.7 & 1.2 & 1.0 & 0.9 & 1.4 & 2.0 & 3.5  \\[2mm]
\hline\hline
\end{tabular}
\end{center}
\end{table}
%


\section{Conclusions}

 In this paper, we have examined the prospect of measuring the Higgs
boson mass using the MAOS neutrino momenta in the Higgs decay
$H\rightarrow WW\rightarrow l\nu l\nu$. To optimize the efficiency
of the method, we have employed an event selection to combine the
well known dilepton azimuthal angle cut, $\Delta \Phi_{ll}< \Delta
\Phi_{ll}^{\rm cut}$, with an $M_{T2}$ cut selecting only the events
with $M_{T2}> M_{T2}^{\rm cut}$. This $M_{T2}$ cut enhances the
efficiency  of the MAOS momentum approximation to the true neutrino
momentum, and the signal to background ratio also. Under such
selection, the MAOS Higgs mass distribution constructed with the
measured charged lepton momenta and the MAOS neutrino momenta shows
a clear peak at the true Higgs boson mass. Likelihood fit analysis
for the MAOS Higgs mass distribution suggests that it can provide a
precise determination of the Higgs boson mass, and also be useful
for the discovery or exclusion of the Higgs boson in certain mass
range.

Our analysis can be improved in many respects, e.g. with more
extensive study of backgrounds, with more complete detector
simulation, and with multi-dimensional fitting including other
observables. At the moment, it is not straightforward to compare our
results with those of Ref.~\cite{Aad:2009wy} providing a more
complete analysis using the transverse mass variable $M_T^{\rm
approx}=M_T(m_{\nu\nu}=m_{ll})$ (together with $\Delta\Phi_{ll}$ and
${\bf p}_T^{WW}$) instead of the MAOS Higgs mass $m_H^{\rm maos}$,
where $M_T^2=m_{ll}^2+m_{\nu\nu}^2+2(E_T^{ll}E_T^{\nu\nu}-{\bf
p}_T^{ll}\cdot{\bf p}_T^{\nu\nu})$ denotes the transverse mass of
$WW\rightarrow l\nu l\nu$. On the other hand, it is relatively easy
to compare our analysis with \cite{Barr:2009mx} which discusses the
Higgs mass determination with $M_T^{\rm true}= M_T(m_{\nu\nu}=0)$ in
a simple context including only the dominant background
$q\bar{q}\rightarrow WW$ without taking into account the detector
effects. Our results appears to be  better than (or comparable to)
those of \cite{Barr:2009mx}. This indicates that incorporating the
collider variable $m_H^{\rm maos}$ might  improve significantly the
accuracy of the Higgs mass measurement, as well as the significance
of the Higgs boson discovery or exclusion.

\vspace{-0.2cm}
\subsection*{Acknowledgements}
\vspace{-0.3cm} We thank W. S. Cho for useful discussions. KC and
CBP are supported by the KRF grants funded by the Korean Government
(KRF-2007-341-C00010 and KRF-2008-314-C00064), KOSEF grant funded by
the Korean Government (No. 2009-0080844), and the BK21 project by
the Korean Government. SC is supported by the KRF grant
(KRF-2006-331-C00072) and the BK21 project.\noindent


%

\end{document}